\documentclass[aps,groupeaddress,prd,english,floatfix,preprint,nofootinbib]{revtex4-1}
\usepackage{epsfig}
\usepackage{amsmath}
\usepackage{slashed}
\usepackage{subfigure,graphicx}
\usepackage{array}
\usepackage{color}

\usepackage{epsfig}
\usepackage{graphicx}

\def \Lag{\mathcal L}

\newcommand{\beq}{\begin{equation}}
\newcommand{\eeq}{\end{equation}}
\newcommand{\beqa}{\begin{eqnarray}}
\newcommand{\eeqa}{\end{eqnarray}}
\newcommand{\bwt}{\begin{widetext}}
\newcommand{\ewt}{\end{widetext}}
\newcommand*{\ovl}{\overline}
\newcommand{\p}{\partial}

\newcommand*{\ra}{\rightarrow}

\newcommand*{\lam}{\lambda}

\newcommand{\nonr}{\nonumber}
\newcommand{\sigv}{\left< \sigma v\right>}
\newcommand{\tdec}{$T_{\mathrm{dec}}\,$}

\begin{document}
\title{ Renormalization Group Study of the Minimal Majoronic Dark Radiation and Dark Matter Model}

%\author[a,1]{We-Fu Chang,\note{Corresponding author.}}
%\author[b]{John N. Ng}

%\affiliation[a]{Department of Physics, National Tsing Hua University,\\ Hsin Chu 300, Taiwan}
%\affiliation[b]{Theory Group, TRIUMF,\\ 4004 Wesbrook Mall, Vancouver BC V6T 2A3, Canada}

% e-mail addresses: one for each author, in the same order as the authors
%\emailAdd{wfchang@phys.nthu.edu.tw}
%\emailAdd{misery@triumf.ca}

\author{We-Fu Chang}
\affiliation{Department of Physics, National Tsing Hua University,Hsin Chu 300, Taiwan}
\author{John N. Ng}
\affiliation{Theory Group, TRIUMF, 4004 Wesbrook Mall, Vancouver BC V6T 2A3, Canada}
%\emailAdd{wfchang@phys.nthu.edu.tw}
%\emailAdd{misery@triumf.ca}
\date{\today}

\begin{abstract}
%\abstract{
We study the 1-loop renormalization group equation running in the simplest singlet Majoron model constructed by us earlier to accommodate the dark radiation and dark matter content in the universe. A  comprehensive numerical study was performed to explore the whole model parameter space. A smaller  effective number of neutrinos $\triangle N_{eff}\sim 0.05$, or a   Majoron decoupling temperature  higher than the charm quark mass, is preferred.
We found that a heavy scalar dark matter, $\rho$, of mass $1.5-4$ TeV is required by the stability of the scalar potential
and an operational type-I see-saw mechanism for neutrino masses. A neutral scalar, $S$, of mass in the $10-100$ GeV range
and its mixing with the standard model Higgs as large as $0.1$ is also predicted. The dominant decay modes are  $S$ into $b\bar{b}$ and/or $\omega\omega$. A sensitive search will come from rare $Z$ decays
via the chain $Z\ra S+ f\bar{f}$, where $f$ is a Standard Model fermion, followed by $S$ into a pair
of Majoron and/or b-quarks.
The interesting consequences of dark matter bound state due to the sizable  $S\rho \rho$-coupling
are discussed as well. In particular,  shower-like events with an apparent neutrino energy at $M_\rho$ could contribute to the
observed effective neutrino flux in underground neutrino detectors such as IceCube.
%}
\end{abstract}

\maketitle
\section{Introduction}

In two previous studies~ \cite{CN},\cite{CNW} we have constructed extensions
of the singlet Majoron model \cite{CMP}\,\cite{SV} with the motivation of accommodating possible new relativistic degree of freedom commonly known as
dark radiation (DR) in cosmological models and also to provide a viable dark matter (DM) candidate. The Majoron which is the Goldstone boson from the spontaneous breaking of the global $U(1)_{\ell}$ lepton symmetry is identified as the DR. This breaking is facilitated by a Standard Model (SM) singlet carrying lepton number of two units. We then add a non-Higgs singlet complex scalar with lepton number $\ell=1$. After symmetry breaking  a stable  scalar DM is obtained. This model preserves the simplicity of the Majoron model and connects the Type I seesaw mechanism to the dark sector which consists of dark matter and perhaps dark radiation. Moreover, the main motivation was to study the physics consequences of identifying the Majoron as
dark radiation. In particular if the decoupling temperature \tdec  is at the muon mass, $m_\mu$ it will give a contribution to the effective relativistic degree of freedom $\Delta N_{\mathrm{eff}}=.39$ which is a sweet spot pointed out in \cite{Weinberg} yielding $N_{\mathrm{eff}}=3.44$. This is higher but not inconsistent with the 2015 Planck result
of $N_{\mathrm{eff}}=3.15 \pm .23$ at $1\sigma$ C.L.\cite{Pl2015}. When other data such as South Pole Telescope and Atacama Cosmology Telescope results are included the central value is higher.
 Although the Planck 2015 data is consistent with the SM prediction, but the statistical significance to rule out DR is still very poor. For example, $\triangle N_{eff} = .33 $ is still consistent with the Planck data at $1\sigma$ C.L. Since a massless Majoron is automatically built in in this model which implements a spontaneously-broken global $U(1)_l$ and it always contributes to the dark radiation. The relevant question is to determine how much can it contribute to $\triangle N_{eff}$.
 This prompted us to reexamine the Majoron dark radiation model by allowing \tdec  to be higher than $m_\mu$ which will reduce $\Delta N_{\mathrm{eff}}$. For example
for \tdec  around 2 GeV $\Delta N_{\mathrm{eff}}=.05$ since now more degree of freedom contributes to the energy density. A consequence pointed out in \cite{CN} and \cite{Weinberg} when taking \tdec at $m_\mu$ gives rise to a scalar of mass less than a few GeV. We find that increasing \tdec will raise the mass of this scalar to the tens of GeV range. This will in turn change the impact on Higgs boson decays since this scalar in general mixes with the SM Higgs boson.

In a recent study \cite{NP} the impact of vacuum stability on the singlet Majoron model with high scale Type I seesaw model for neutrino masses was investigated. Besides the high seesaw scale the mechanism also requires Yukawa couplings of the righthanded Majorana neutrinos to active neutrinos via the Higgs field. Spontaneous electroweak symmetry breaking gives rise to a Dirac mass term. This is the second crucial ingredient of type I  seesaw mechanism. Hence,
 for it to work the electroweak vacuum must be stable for when lepton number breaking occurs. In other words the seesaw scale must be lower than the energy, $\mu_{VS}^{SM}$, where the electroweak vacuum becomes unstable which is known to be around  $\mu_{VS}^{SM}\simeq 10^{10}-10^{12}$ GeV \cite{Vs}\cite{IRS}\cite{Buttazzo:2013uya}. For the Majoron model lepton number is spontaneously broken and hence the stability of the singlet scalar that breaks this symmetry must also be taken into account. It was found that the stability of the SM can be extended to the  GUT
 scale without invoking metastability.  See \cite{axionDR} for a similar  discussion on the vacuum stability by identifying the axion as DR.

 In this paper we study how RG considerations impact the  parameters of dark matter and dark radiation sector of the Majoron model. We
calculate the one loop beta function for the renormalization  group running
of the all the relevant parameters of the Majoron dark radiation model of \cite{CN}. We find that  the stability of the scalar vacua has very important effect on
the parameters of the theory. In particular the dark matter candidate $\rho$ will have a mass in the range of the lepton number violation scale; i.e. in the
 several TeV range. This is vastly different from the usual studies which did not  take into
 account scalar vacua stability. We find that in doing so can lead to interesting astroparticle
 physics consequences. Since the DM is heavy and there is a light scalar in the spectrum, bound state of DM can be formed if the triple scalar coupling is strong enough.
 We show that this can indeed take place in a large region of the model parameter space. We speculate that DM annihilation into two Goldstone bosons will be enhanced. One would then expect a Goldstone component in the high
 energy cosmic ray spectrum.

 We organize the paper as follows. In section two we give a summary of the model and the RGEs of the relevant parameters. It is sufficient to
 use the 1-loop result for the beyond  SM physics. This is followed by details of the numerical study of the model including the solutions of the RGEs.
 In section IV we discuss the phenomenological consequences of the results we obtained. Finally we conclude in section V.

\section{The Model}

A singlet Higgs field $S$ which carries lepton number $\ell=2$ and a non-Higgssed scalar field $\Phi$ with $\ell=1$ are added to the particle contents of the SM.
Realistic  implementation of the Type-I seesaw mechanism will require adding at least two singlet Majorana right-handed
neutrinos $N_{Ri}, i=1,2$, to the SM. Since the details of the neutrino physics such as active neutrino masses and oscillations are not relevant to this study and we can just take one righthanded neutrino for simplicity without affecting the physics we are interested in. Extending to the realistic case of two or more righthanded neutrinos is straightforward. Also,the SM Higgs field is denoted by $H$.

Due to the $U(1)_l$ symmetry, $\Phi$ will not have a trilinear coupling with
$H$;thus, it will not contribute to the Majorana mass of $N_{R}$. Its Dirac mass type of couplings to the active neutrinos
are also forbidden since it is an $SU(2)$ singlet. Therefore, much of the Majoron model is not changed
and its simplicity is retained.

The most general scalar lagrangian is given as
\begin{align}
\label{eq:lscalar}
{\Lag}_{scalar}&= (D_\mu H)^\dagger (D^\mu H)  + (\p_\mu \Phi)^\dagger (\p^\mu \Phi)+ (\p_\mu S)^\dagger (\p^\mu S) -V(H,S,\Phi)\,,  \nonr \\
V(H,S,\Phi)&= -\mu^2 H^\dagger H -\mu_s^{2} S^\dagger S
+m_{\Phi}^2 \Phi^\dagger \Phi  +\lambda_H (H^\dagger H)^2+ \lambda_{\Phi} (\Phi^\dagger \Phi)^2\nonr \\
&{\phantom{=}} +\lambda_s (S^\dagger S)^2 +\lambda _{SH} (S^\dagger S)(H^\dagger H)  +\lambda_{\Phi H}(\Phi^\dagger \Phi)(H^\dagger H) \nonr \\
&{\phantom{=}}+ \lambda_{\Phi S}(S^\dagger S)(\Phi^\dagger \Phi)+\frac{\kappa}{\sqrt 2}\left[ (\Phi^\dagger)^2 S + S^\dagger \Phi^2\right]\, ,
\end{align}
and we take $\kappa$ to be real and $m_{\Phi}^2 >0$ so that $\langle \Phi\rangle=0$. Using the usual linear representation of scalar fields we expand them as follows
\begin{align}
\Phi&= \frac{1}{\sqrt 2} (\rho + i\chi)\,, \nonr \\
S&= \frac{1}{\sqrt 2} (v_s +s + i\omega)\,,
\end{align}
and use the U-gauge for the Higgs
\beq
H=\begin{pmatrix}0\\ \frac{v_H+h}{\sqrt 2}
\end{pmatrix}\,,\,\,\mbox{$v_H= 246$ GeV}.
\eeq
The physical fields are $ \hat {S} =(h,s,\rho,\chi)$ and $\omega$ is the massless Goldstone boson named the  Majoron. With this one obtains the scalar mass matrix  squared:
\beq
\label{eq:smass}
\begin{split}
 M^2& = \\
& \begin{pmatrix} 2\lambda_H v_H^2 & \lambda_{SH}v_H v_s &0 &0 \\
\lambda_{SH} v_H v_s & 2\lambda_s v_s^{2} &0 &0 \\
0& 0& m^2_{\Phi} +\frac{1}{2}\lambda_{\Phi H}v_H^2+\frac{1}{2}\lambda_{\Phi S}v_{s}^2+ \kappa v_s &0 \\
0 & 0 &0 & m_{\Phi}^2 +\frac{1}{2}\lambda_{\Phi H}v_H^2 +\frac{1}{2}\lambda_{\Phi S}v_{s}^2- \kappa v_s
\end{pmatrix}\,.
\end{split}
\eeq
Note that $\kappa$ splits the degeneracy of the $\rho$ and $\chi$ masses and we require $m^2_{\Phi}>|\kappa v_s|-\frac{1}{2}(\lambda_{\Phi h}v_H^2+\lambda_{\Phi S}v_s^2)$.

We take $\rho$ to be the DM and its stability is guaranteed by $Z_2$ dark parity which remains after
spontaneous symmetry breaking of $U(1)_\ell$ \cite{CN}.

In terms of component fields the scalar potential becomes
\beq
\label{eq:scalarpot}
\begin{split}
V=& \frac{1}{2} \tilde{\hat{S}}M^2 \hat{S} +\lambda_H v_H h^3 +\frac{1}{4}\lambda_H h^4 + \lambda_s v_s s^3 +\lambda_s v_s \omega^2 s +\frac{1}{4}\lambda_s( s^4 +\omega^4)+ \frac{1}{2} \lambda_s \omega^2 s^2\\
&+\frac{1}{4}\lambda_{\Phi}(\rho^4 +\chi^4 +2\rho^2\chi^2) +\frac{1}{2}\lambda_{SH} v_s s h^2 +\frac{1}{2}\lambda_{SH}v_H(s^2 +\omega^2)h +\frac{1}{4}\lambda_{SH}(s^2+\omega^2)h^2\\
&+ \frac{1}{2}\lambda_{\Phi H} v_H (\rho^2+\chi^2)h+ \frac{1}{4}\lambda_{\Phi H}(\rho^2 +\chi^2)h^2 +\frac{1}{4}\lambda_{\Phi S}\left (s^2\rho^2+s^2\chi^2+\omega^2\rho^2+\omega^2\chi^2\right )\\
&+\frac{1}{2}\bar{\kappa} s\,\rho^2 +\frac{1}{2}(\bar{\kappa} -2\kappa)s\,\chi^2  + \kappa \rho\chi \omega \,,
\end{split}
\eeq
where $\bar{\kappa}\equiv\lambda_{\Phi S} v_s +\kappa$.

It is clear that $(h,s)$ are not yet mass eigenstates denoted by $(h_1,h_2)$.
They are related by the usual rotation:
\begin{equation}
\begin{pmatrix}
h_1 \\
h_2
\end{pmatrix}
=
\begin{pmatrix}
\cos\theta & -\sin\theta \\
\sin\theta & \cos\theta
\end{pmatrix}
\begin{pmatrix}
h \\
s
\end{pmatrix} \,,
\end{equation}
with the mixing angle $\theta$ given by
\begin{equation}
\label{eq:thmix}
\tan 2\theta = \frac{\lam_{HS}v_H v_S}{\lam_S v_S^2 - \lambda_H v_H^2} \,.
\end{equation}
We shall identify $h_1 \equiv H$ as the SM Higgs which has a mass of 125~GeV.
Note that in the small mixing limit, $M_{H}^2 \approx 2\lambda_H v_H^2$, $m_2^2 \approx 2\lam_S v_S^2$, $h_1\approx H$, and $h_2\approx S$.

The Lagrangian responsible for the seesaw mechanism is given by
\beq
-{\Lag}_\ell= y_\nu \ovl{L}_L\tilde {H} N_R + Y_S \ovl{N_R^c} N_R S + h.c.
\eeq
where $L= (n_L, e_L)^T$ is the SM lepton doublet and $\tilde {H}=i\sigma_2 H^*$. After symmetry breaking we get
\beq
-{\Lag}_\ell=\frac{y_\nu v}{\sqrt 2}\ovl{n_L}N_R + \frac{Y_S v_s}{\sqrt 2} \ovl{N^c_{R}} N_R + \frac{y_\nu}{\sqrt 2} \ovl{n_L} N_R h + \frac{Y_S}{\sqrt 2} (s+i\omega)\ovl{N^c_{R}}N_R +h.c.
\eeq

We wish to identify $\omega$ as the DR and the amount it contributes to $\Delta N_{\mathrm{eff}}$ depends on when it decouples from the thermal bath.
In particular we are interested at decoupling temperatures around QCD phase transition, charm mass and tau mass; then $\Delta N_{\mathrm{eff}} = 0.055, 0.0451, 0.0423 $ respectively. The effective Lagrangian for $\omega \omega \ra f\bar{f}$ is given by
\beq
{\Lag}_{f\omega} \sim -\frac{\lambda_{HS}m_{f}}{M_H^2 M_s^2}\bar{f} f \partial_\mu\omega \partial^\mu\omega
\eeq
where $f$ denotes the SM fermion in the thermal bath.
The rate of $f\bar{f} \leftrightarrow \omega\omega$ is estimated to be
\beq
\Gamma(f\bar{f} \leftrightarrow \omega\omega) \sim \frac{\lambda_{HS}^2 m_f^2}{ M_H^4 M_S^4} \times T_{\mathrm{dec}}^7\times N_c^f
\eeq
where $N_{c}^f$ is the color of $f$. Apparently the specie, which will be just denoted as $f$,  with the largest values of $ (m_f^2 \times N_c^f)$ dominates the process which is proportional to $m_{eff}^2(T_{dec}) \equiv \sum_{m_f<T_{dec}} N_c^f m_f^2$.
In order for $\omega$ to play the role of DR the collision rate of $\omega$ into a pair of fermions must be approximately the Hubble expansion rate at \tdec, thus;
\beq
\label{eq:dec}
\frac{N_c\lambda_{HS}^2m_{eff}^2 T_{\mathrm{dec}}^5 M_{Pl}}{M_H^4 M_s^4} \approx 1\,.
\eeq
For \tdec  at around $m_\tau$,  both charm and tau must be considered and we have  $N_c m_f^2 \ra N_c m_c^2+m_\tau^2$; otherwise $m_f$ is the mass of fermion nearest to the \tdec. In Eq.(\ref{eq:dec}) the parameter
$\lambda_{HS}$ is controlled by the mixing of the scalar $S$ with the Higgs. It is expected to be small. Eq.(\ref{eq:dec}) shows that $M_s$ is in the 100 GeV range
if \tdec  is around the charm mass. This is to be compared to decoupling at $m_\mu$ which leads to a light scalar of mass less than a few GeV. Details for the latter case can be found in \cite{CN}.

\begin{figure}[ht]
    \centering
    \subfigure[]{\includegraphics[width=6cm]{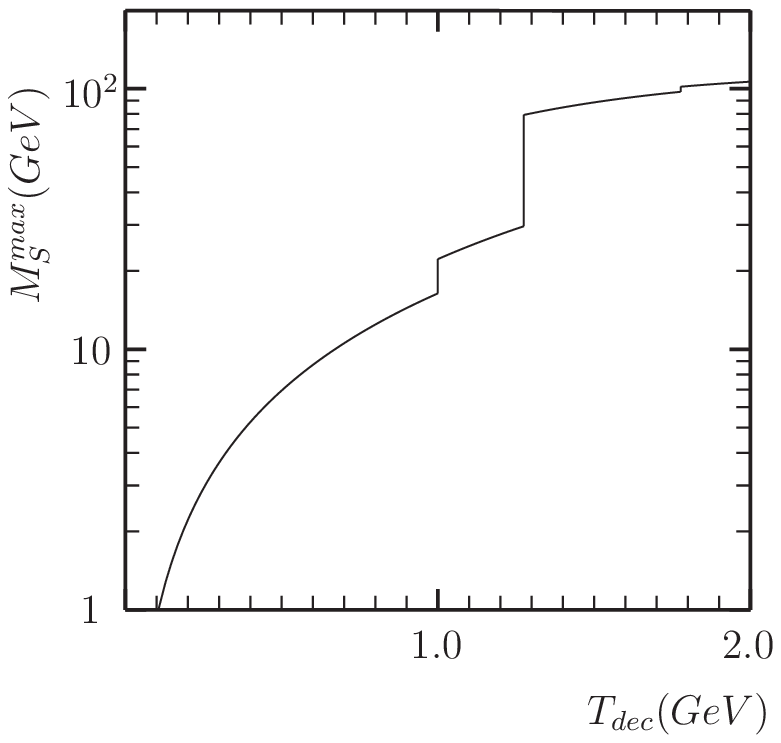}}% \label{fig:RGE4HS}}%
    \qquad
     \subfigure[]{\includegraphics[width=6.5cm]{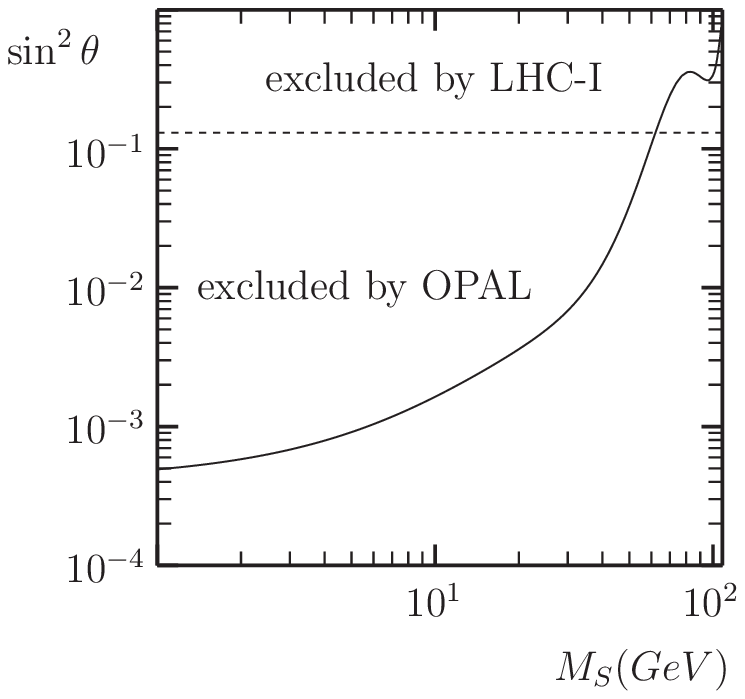}}% \label{fig:RGE4PH}}
    \caption{(a) Upper bound of $M_S$ from the SM Higgs invisible decay width $\Gamma(H\ra \omega\omega)<0.8$ MeV. Here we take into account of the contribution from the three light quarks at $T_{dec}\sim 1$ GeV, and we set $m_{u,d}\sim 4$MeV, $m_s=95$MeV, and $m_c=1.275$ GeV.
    (b) The 2 $\sigma$ experimental upper bound on $\sin^2\theta$ v.s. $M_S$ from OPAL\cite{OPAL} and LHC run-I\cite{LHC_mu}(dash line).}%
    \label{fig:MS_max_mixing}%
\end{figure}

 For a given \tdec the largest $M_S$ can be estimated by considering the Higgs invisible decay width, which is experimentally given by $\Gamma_H^{inv}<0.8$ MeV \cite{Giardino:2013bma, Djouadi:2005gi}.
 Since the Majoron is massless,  the SM Higgs can always decay into two Majorons and this width is denoted by $\Gamma_{\omega\omega}$. There may be other invisible modes available; thus $\Gamma_{\omega\omega}\leq
 \Gamma_H^{inv}.$ Moreover, $\Gamma_{\omega\omega}$ is given by

  \beq
  \label{eq:2omega}
  \Gamma_{\omega\omega} = \frac{1}{32\pi}\frac{\sin^2\theta M_H^3}{v_S^2}
  \eeq
  in terms of the physical mass of the SM Higgs.
  Using the relations between  $\sin\theta$ and $v_S$ and the mass eigenvalues plus the decoupling condition, Eq.(\ref{eq:dec}), we can rewrite the above as
 \beq
  {M_S^4 \over (M_H^2-M_S^2)^2} \leq \cos^2{\theta} {32\pi m_{eff}^2 T_{\mathrm{dec}}^5 M_{pl} \over v_H^2 M_{H}^7} \Gamma_{H}^{inv} \leq {32\pi m_{eff}^2 T_{\mathrm{dec}}^5 M_{pl} \over v_H^2 M_{H}^7} \times (0.8 \mbox{MeV})\,.
  \eeq
  From the above inequality, the upper bond of $M_S$ can be easily solved analytically. We should just denote the solution  as $M_S^{max}(T_{\mathrm{dec}})$, shown in Fig.\ref{fig:MS_max_mixing}(a), since the explicit form is not important.

A direct search for the light neutral scalar denoted by $S$ here, at OPAL\cite{OPAL} yields an upper limit of the size of mixing between $H$ and $S$.
On the other hand, the mixing will modify the SM Higgs coupling to anything by a $\cos\theta$ factor.
At the LHC, the signal strength $ \mu^f_i$ for a specific production and decay channel $i\ra H\ra f$
is defined as
\beq
 \mu^f_i \equiv { \sigma_i \cdot BR^f \over (\sigma_i)_{SM}\cdot (BR^f)_{SM}}\,.
\eeq
If  the SM Higgs invisible decay width $\Gamma_H^{inv} \ll \Gamma_H^{SM}$, which is the case in our numerical study, then $BR^f \simeq (BR^f)_{SM}$  and $\mu^f_i \simeq \cos^2\theta$  is predicted in our model.
The best-fit  of signal strength $\mu=1.1 \pm 0.11$ is given  in a recent ATLAS and CMS combined global analysis on all production process and decay channels  with data taken at $\sqrt{s}=7$ and $8$ TeV\cite{LHC_mu}.
This indirect bound amounts to $\sin^2\theta <0.13$ at 2 $\sigma$ level which has also been implemented for $M_S>60$ GeV in our numerical study, Fig.\ref{fig:MS_max_mixing}(b).
 This latest bound is derived from LHC Run-1 data only, and the expected sensitivity of Run-2 will be discussed in the phenomenology section later. 

Although many of the parameters in the scalar potential Eq.(\ref{eq:scalarpot}) are unknown, we can gain some information by demanding that the stability of the scalar sector in the appropriate range.
The stability of the electroweak vacuum govern by the sign of $\lambda$ is well known to be at best metastable \cite{Vs,IRS} for the SM. On the other hand singlet scalars
generally helps to stabilize the electroweak vacuum. In our model we also require that both $S$ and $\Phi$ should have stable potentials for consistency reasons.
The scales of stability are given by the RG running of the parameters. The RGEs for the SM couplings are easily found in the literature and we will not
repeat them. The relevant RGEs for the new parameters calculated with 1-loop $\beta$ functions are given below:
\begin{subequations}
\label{eq:RGE}
\beqa
\label{eq:RGEa}
16\pi^2\frac{d\lambda_H}{dt}&=&12\lambda_H^2 + 6\lambda_H y_t^2 -3y_t^4 -\frac{3}{2}\lambda_H (3g_2^2 + g_1^2)+\frac{3}{16}\left[(g_1^2+g_2^2)^2+2g_2^4\right] \nonr \\
&&+\frac{1}{2}(\lambda_{HS}^2+\lambda_{\Phi H}^2)\,,  \\
16\pi^2 \frac{d\lambda_\Phi}{dt}&= &10\lambda_\Phi ^2 + \lambda_{\Phi H}^2 +\frac{1}{2} \lambda_{\Phi S}^2\,, \\
16\pi^2 \frac{d\lambda_S}{dt}&=& 10 \lambda_S^2 +\lambda_{HS}^2 +\frac{1}{2}\lambda_{\Phi S}^2 -8 Y_S^4 + 4Y_S^2 \lambda_S\,,  \\
16\pi^2 \frac{d\lambda_{HS}}{dt}&=& 2\lambda_{HS}^2 +\lambda_{HS}(6\lambda_H +4\lambda_S)+2\lambda_{HS}Y_S^2 -\frac{3}{4}\lambda_{HS}(3g_2^2+g_1^2)\nonr\\
&&+3\lambda_{HS}y_t^2 + \lambda_{\Phi S} \lambda_{\Phi H}\,, \\
16\pi^2 \frac{d\lambda_{\Phi H}}{dt}& =& 2\lambda_{\Phi H}^2 +\lambda_{\Phi H} (6\lambda_H +4\lambda_\Phi)- \frac{3}{4} \lambda_{\Phi H} (3g_2^2 +g_1^2)\nonr\\
 && + 3\lambda_{\Phi H} y_t^2  +  \lambda_{\Phi S} \lambda_{HS}\,, \\
16\pi^2 \frac{d\lambda_{\Phi S}}{dt}&=& 2\lambda_{\Phi S}^2 + 4\lambda_{\Phi S}(\lambda_{\Phi}+\lambda_S) +2\lambda_{\Phi S}Y_S^2
+ 2 \lambda_{\Phi H} \lambda_{HS}\,,  \\
16\pi^2 \frac{d Y_S}{dt}&=& 3Y_S^3\,, \\
16\pi^2 \frac{ d\kappa}{dt} &=&\kappa \left( 2\lambda_{\Phi S} +2\lambda_\Phi + Y_S^2 \right)\,,
\eeqa
\end{subequations}
where $y_t,g_2,g_1$ are the t-quark Yukawa coupling, $SU(2)$ and $U(1)_Y$ gauge couplings respectively and $t\equiv \ln {\frac{Q^2}{Q_0^2}}$ with $Q_0$ an arbitrary renormalization point. We have omitted all light fermion Yukawa couplings including
those of the active neutrinos since they are all very small. The running of the $y_\nu$'s can be shown to be unimportant for us \cite{NP}. For stability we require  $\lambda_i >0$ and $\lambda_{ij}> -2\sqrt{\lambda_i \lambda_j}$ where $i,j = H, S, \Phi$. The RGE's for these couplings by themselves are not sufficient to determine whether any of them will turn negative at high enough energies. We need boundary conditions at some lower energies.
Since $M_S <100$ GeV, we choose this scale to be $m_Z$. The values of the couplings at this scale will be given by numerical scan that has to satisfy other constraints we impose on the model. Details are given in the next section.

One important input comes from DM considerations. Our addition to the minimal Majoron model produces a WIMP DM candidate. Due to a $Z_2$ dark parity the lighter of $\rho$ and $\chi$ will be the DM. Without loss of generality we take  that to be $\rho$ and their masses are split by $M_\chi^2-M_\rho^2= -2\kappa v_s$ which is not small as we shall see later. The relic density of $\rho$ can be calculated by evaluating the rate $\rho \rho$ annihilating into a pair of SM particles or new scalars $ss, \omega\omega, Hs$. The complete list is given in \cite{CN}. The controlling quantity that determines the relic density of WIMP DM is the thermally averaged annihilation cross sections $\sigv$. The relevant ones are given below
\begin{subequations}
\label{eq:sigv}
\beqa
( \sigma v )_{ss} &=& \frac{1}{64\pi}\frac{\sqrt{1-x_s}}{M_\rho ^2}\left[ q_{SS} +{g_{\rho H} \lambda_{HSS}\over M_\rho^2(4-x_H) }
+{g_{\rho S} \lambda_{SSS}\over M_\rho^2(4-x_S) }-\frac{2g_{\rho S}^2}{M_{\rho}^2(2-x_s)}\right]^2 \,,  \\
( \sigma v )_{HH} &=& \frac{1}{64\pi}\frac{\sqrt{1-x_H }}{M_\rho ^2}\left[ q_{HH}+  \frac{g_{\rho H} \lambda_{HHH}}{M_\rho^2(4-x_H)}+ \frac{g_{\rho S} \lambda_{SHH}}{M_\rho^2(4-x_S)}-\frac{2 g_{\rho H}^2 }{M_\rho^2 (2-x_H)}\right]^2\,, \\
( \sigma v )_{Hs} &=&\frac{1}{32\pi}\frac{\Delta}{M_{\rho}^2}\left[ q_{HS}+ \frac{g_{\rho H} \lambda_{SHH} }{M_{\rho}^2(4-x_H)} + \frac{g_{\rho S} \lambda_{HSS} }{M_{\rho}^2(4-x_S)}
-\frac{4 g_{\rho H} g_{\rho S}}{M_{\rho}^2(4-x_H -x_s)}\right]^2\,,  \\
( \sigma v )_{\omega\omega}&=& \frac{1}{64\pi M_{\rho}^2}\left[\frac{g_{\rho H} g_{\omega H}}{M_\rho^2 (4-x_H)}+\frac{g_{\rho S} g_{\omega S}}{M_\rho^2 (4-x_S)}+\lambda_{\Phi S}-\frac{2\kappa ^2}{M_\rho ^2(1+x_\chi)}\right]^2\,,  \\
( \sigma v )_{WW} &=&\frac{1}{8\pi}\frac{\lambda_{\Phi H}^2}{M_\rho ^2}\sqrt{1-x_W}\left[4-4x_W +3x_{W}^2\right] \left[\frac{c_\theta^2}{(4-x_H)} +\frac{s_\theta^2}{(4-x_S)} \right]^2 \,, \\
( \sigma v )_{ZZ} &=& \frac{1}{16\pi}\frac{\lambda_{\Phi H}^2}{M_\rho ^2}\sqrt{1-x_Z}\left[4-4x_Z +3x_{Z}^2\right] \left[\frac{c_\theta^2}{(4-x_H)} +\frac{s_\theta^2}{(4-x_S)} \right]^2 \,, \\
( \sigma v )_{f\bar{f}}&=& \frac{N_c}{4\pi}\frac{\lambda_{\Phi H}^2 x_f}{M_\rho ^2}(1-x_f)^{\frac{3}{2}}
\left[\frac{c_\theta^2}{(4-x_H)} +\frac{s_\theta^2}{(4-x_S)} \right]^2\,,
\end{eqnarray}
\end{subequations}
where $\Delta^2\equiv 1+\frac{1}{16}x^2_{H}+\frac{1}{16}x^2_{s}-\frac{1}{8}x_H x_s -\frac{1}{2}x_H-\frac{1}{2}x_s$, $x_i\equiv\frac{M_i ^2}{M_\rho ^2}$ for $i=W,Z,H,f,S,\chi$, and the subscripts denote the final state. The couplings in the scalar mass eigenstates are given as
\beqa
q_{SS}&=&\lambda_{\Phi S} c^2_\theta +\lambda_{\Phi H} s^2_\theta\,,\;
q_{HH}=\lambda_{\Phi S} s^2_\theta +\lambda_{\Phi H} c^2_\theta\,,\;
q_{HS}=(\lambda_{\Phi H}-\lambda_{\Phi S}) c_\theta s_\theta\,,\nonr\\
g_{\rho S}&=& \bar{\kappa} c_\theta +\lambda_{\Phi H} v_H s_\theta\,,\;
g_{\rho H}= -\bar{\kappa} s_\theta +\lambda_{\Phi H} v_H c_\theta\,,\nonr\\
g_{\omega H}&=& \lambda_{S H} v_H c_\theta -2 \lambda_{S} v_S s_\theta\,,\;
g_{\omega S}= \lambda_{S H} v_H s_\theta + 2 \lambda_{S} v_S c_\theta\,,\nonr\\
\lambda_{HHH}&=& 6\lambda_H v_H c_\theta^3 -6\lambda_S v_S s_\theta^3 + 3\lambda_{SH}  s_\theta c_\theta (v_H s_\theta -v_S c_\theta)\,,\nonr\\
\lambda_{SSS}&=& 6\lambda_H v_H s_\theta^3 +6\lambda_S v_S c_\theta^3 + 3\lambda_{SH}  s_\theta c_\theta (v_S s_\theta + v_H c_\theta)\,,\nonr\\
\lambda_{SHH}&=& 6 s_\theta c_\theta( \lambda_H v_H c_\theta +\lambda_S v_S s_\theta ) + \lambda_{SH} v_S( c_\theta^3-2 s_\theta^2 c_\theta)
+ \lambda_{SH} v_H ( s_\theta^3-2 s_\theta c_\theta^2)\,,\nonr\\
\lambda_{HSS}&=& 6 s_\theta c_\theta( \lambda_H v_H s_\theta -\lambda_S v_S c_\theta ) + \lambda_{SH} v_S(-s_\theta^3+2 s_\theta c_\theta^2)
+ \lambda_{SH} v_H ( c_\theta^3-2 s_\theta^2 c_\theta)\,.
\label{eq:coupling_mass_basis}
\eeqa
At high temperatures these will give $\sigv$ and for the correct relic abundance the total $\sigv$ should be $\sim 3\times 10^{-26} {\mathrm{cm}}^3/s$.
A numerical scan is performed as described in the next section to obtain possible values of unknown couplings.

 Next we discuss the constraint imposed by the limits from direct DM detection since there is no convincing signals yet. In our model the scattering of $\rho$
off the nucleus of the detector will deposit energy.  The scattering $\rho +n \ra \rho +n$ where $n$ denotes a nucleon proceeds via the t-channel exchange of $H$ and $S$.
 It is often to parameterize the SM Higgs-nucleon-nucleon coupling by $\eta g_2 M_n/(2 M_W)$ \cite{HiggsHunter} where $M_n$ is the nucleon mass and $\eta$ is a parameter represents the uncertainty in the coupling. In the interaction basis the $h\,n\,n$ and $s\,n\,n$ couplings become $ c_\theta \eta g_2 M_n/(2 M_W)$ and $ s_\theta \eta g_2 M_n/(2 M_W)$ respectively. And the tree-level  cross section in terms of physical masses of $H$ and $S$ is
\beq
\label{eq:DMnucleon}
\sigma_{\rho \mathrm{n}}=\frac{G_F M_{n}^2 \eta^2 m_r^2(n,\rho)}{4\sqrt{2}\pi M_{\rho}^2 M_{H}^2 \lambda_H}\left[\lambda_{\Phi H}\left(c^2_{\theta}+s^2_{\theta}\left(\frac{M_H}{M_S}\right)^2 \right)-s_\theta c_\theta\frac{\bar{\kappa}}{v_H}\left(1-\left(\frac{M_H}{M_S}\right)^2\right)\right]^2\,,
\eeq
where the reduced mass is
\beq
m_r(n,\rho)=\frac{M_\rho M_n}{M_\rho + M_n}\,.
\eeq
 We take $\eta=.3$  which is the value obtained from QCD consideration\cite{HiggsHunter} and ignore  possible isospin breaking effects and the strange quark content in the nucleon. 
These can be incorporated as given in \cite{Crivellin}. As can be seen above direct detection can strongly constrain $\lambda_{\Phi H}$ and
$\frac{\bar{\kappa}}{v_H} s_{2\theta}$.

\section{Numerical study}

\subsection{Scan strategy}
A numerical study of the parameter space is performed as follow. We scan the full parameter space according to the following order:
\begin{itemize}
  \item A value of \tdec is randomly chosen in the range between  $m_\mu$ and $2$ GeV.

  \item Then we randomly pick $M_S \in [(m_K-m_\pi), M_S^{max}(T_{\mathrm{dec}}) ]$. The lower bound is chosen
   to avoid the stringent experimental bound on $K\ra \pi +(\mbox{nothing})$.
   Furthermore, the  phenomenology  of scalars as light as that was discussed in \cite{CN} and we will not repeat it.
    Once \tdec and $M_S$ are fixed, $\lambda_{SH}$ is determined by Eq.(\ref{eq:dec}).

  \item The value of $|\theta|$ is randomly generated  within $ |\theta|< \theta^{max}(M_S)$.\\
    The upper bound $\theta^{max}(M_S)$ is given by the  OPAL direct search for $M_S>1$ GeV\cite{OPAL} and the indirect bound from LHC run-I\cite{LHC_mu} for $M_S> 60$ GeV as discussed in previous section. We only found a few viable solution for $M_S<2$ GeV in our numerical scan, however  we did not exclude this possibility in our study. We set the upper bound of $|\theta|<2\times 10^{-3}$ for $M_S<2$ GeV which comes  mainly from the rare B decays \cite{Anchordoqui:2013bfa} which is much more stringent than the OPAL bound.

  \item The range for  $M_\rho$ is $\in [0.5 \mbox{TeV}, 4 \mbox{TeV}]$.\\
   In our numerical scan we found no solution for DM lighter than $0.5$ TeV. This can be understood as follows.  Since the requirement of RGE improvement of scalar stability will lead to large scalar couplings. Roughly speaking, the larger scalar couplings the larger DM annihilation
  cross section, and hence the smaller relic density.
   So one needs heavier DM to lower this cross section in order to get the relic density in the right ballpark. On the other hand, for the same couplings, the heavier $\rho$ gives higher relic density at freeze out.  Hence; there is an upper bound on $M_\rho$ so that relic density is not so high as to over close the universe. This is conservatively chosen to be 4 TeV.

  \item With the above set of parameters generated  we calculate the following
  \beqa
  \lambda_H &=& {\cos^2\theta M_H^2 +\sin\theta^2 M_S^2 \over 2 v_H^2}\,,\\
   \label{eq:VS_mass_basis}
  v_S &=& - {\sin\theta \cos\theta (M_H^2-m_S^2) \over v_H \lambda_{SH}}\,,\\
    \label{eq:lam_S_mass_basis}
   \lambda_S &=& {\sin^2\theta M_H^2 +\cos\theta^2 M_S^2 \over 2 v_S^2}\,,\\
  y_S &=& \frac{\sqrt{2} M_N}{v_S}\,.
  \eeqa
  With these parameters  we check $\Gamma_H^{inv}$  to make sure the sum of all the invisible decay channels is still
  smaller than the experimental limit. If so this parameter set will be accepted as viable solutions.
\item For $\lambda_{\phi S}$ the range is $ \in [-4\sqrt{\pi \lambda_S},4\pi]$.\\
  The lower bound is from the positivity of the scalar potential and the upper bound is from the perturbativity.

  \item Then $\bar{\kappa}$ is generated in the range $ \in [-v_S, +v_S]$, with $\kappa<0$.\\
  Here, a consistency check is  made so that $\kappa = \bar{\kappa}-\lambda_{\phi S} v_S <0$. This ensures $\rho$ is the DM candidate.

  \item We generate $\lambda_{\phi H}$ in the range$ \in [-4\sqrt{\pi \lambda_H}, 4\pi]$.\\
  With all the above parameters fixed, except $\lambda_\phi$ which has no low energy constraint, we can go on to check whether the relic density and the DM direct search bound \cite{LUX} are both met.
  Otherwise, the process will start over again.

  \item Lastly, we randomly scan $\lambda_\phi \in [0, 4\pi]$. Since there is no known constraint we use the RGE
  to determine its viable value.

  For each $\lambda_\phi$, the whole set RGEs running are carried out.
  The relevant  boundary conditions and parameters we used for RGE running are: $M_Z=91.1876$ GeV, $M_H=125.0$GeV, $M_t=173.0$ GeV\footnote{ At 1-loop level the running of $y_t$ is the SM one. It is well known, see \cite{IRS}, that $\mu_{VS}^{SM}$ is very sensitive to the initial value of $y_t$ (or $m_t$) for RGE running. Since we use the SM as the reference point and all we require is that the lepton number violation scale is below $\mu_{VS}$. The top quark mass uncertainty does not enter to affect our study and conclusions.}, $\alpha_S(M_Z)=0.1184$, $\alpha(M_Z)=1/127.916$, and $\sin^2\theta_W=0.23116$.
  If a Landau pole is   encountered, or $\lambda_{\phi, S}$ become negative, or any of the positivity conditions is violated, i.e $\lambda_{ij}<-2\sqrt{\lambda_i \lambda_j}$, in the stability region of $\lambda_H$ the parameter set is discarded. We denote the scale where the vacuum instability happens as $\mu_{VS}$. If electroweak stability is improved, $\mu_{VS} >\mu_{VS}^{SM(1-loop)}$, the set is considered viable
\footnote{Since we only consider the RGE at the 1-loop level for the new scalars, for consistency, the SM vacuum stability is also determined by the  SM 1-loop RGE. Using the stated input values, the 1-loop SM scalar potential becomes unstable at
 the scale of $\mu_{VS}^{SM(1-loop)}\equiv 1.9 \times 10^{5}$ GeV. Given the exploratory nature of our study we deem this to be sufficient.}.
  If after some large number ($10^5$) of tries without success, the whole set of parameters
    will be discarded and the scan goes to first step again; otherwise we register the parameter set as one viable configuration.

    \end{itemize}
In fact, $M_N$ is also a free parameter in our model. However, our numerical experiment could not find any viable solution for $M_N<0.5$ TeV and the numerical results are not very sensitive to the actual value when $M_N \sim $ few TeV.
Therefore, in our study we just set $M_N =1$ TeV as a benchmark.

\subsection{Overview of the numerical results}
The viable configurations are easy to get if one only requires that the scale of vacuum instability is higher than the SM  one,  $\mu_{VS} >\mu_{VS}^{SM(1-loop)}$.   For later use, we define  $R_{VS} \equiv \log_{10} \frac{\mu_{VS}}{\mu_{VS}^{SM(1-loop)}}$
to quantify how much the improvement of the  vacuum stability scale comparing to the SM case.
To emphasize how the vacuum stability and RGEs affect the parameters in our model, here we focus on those configurations with $R_{VS} >2$ and we have generated 4000 such viable sets of parameters. This choice is arbitrary and we intend it for illustration purpose only.
 We found many configurations with $R_{VS}>2$ and the largest $R_{VS}$ we got is $\sim 11$ using the scan algorithm just stated.
  This is in agreement with the expectation that singlet
scalars or Higgs portal models tend to improve electroweak vacuum stability. With this algorithm and the computing resource at hands, we did not
find the configurations where the scale is pushed all the way to the Planck mass.

\begin{figure}[ht]
    \centering
    \subfigure[]{\includegraphics[width=5cm]{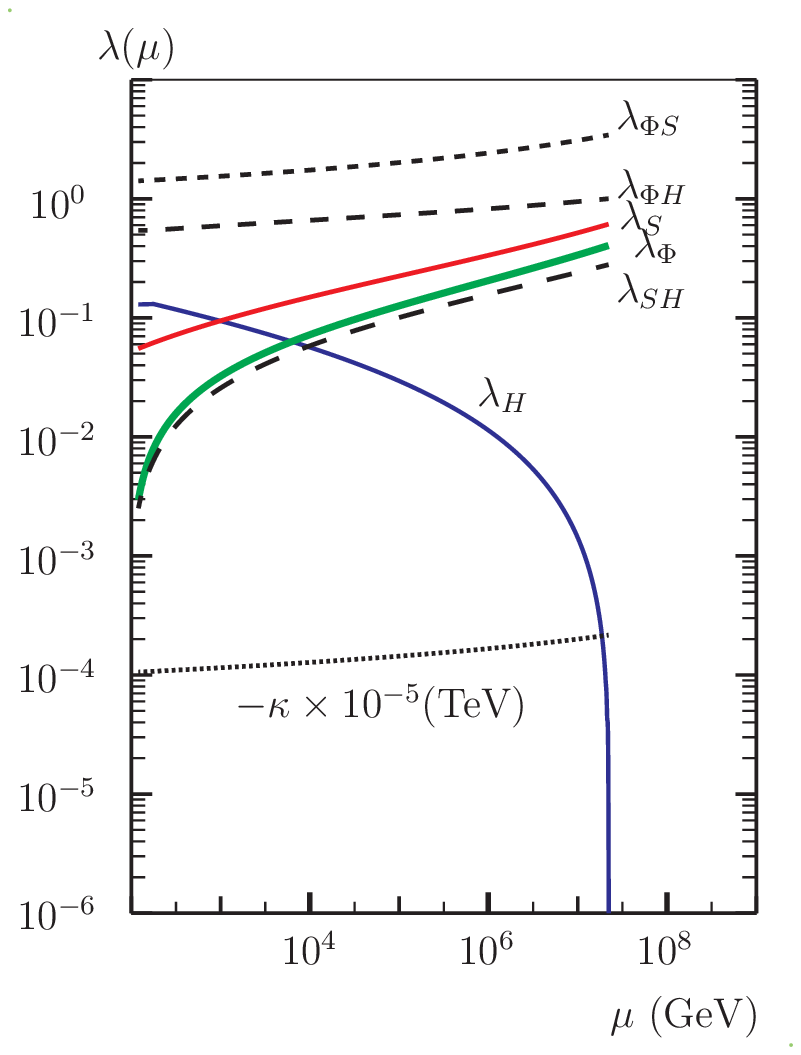}}% \label{fig:RGE4HS}}%
    \qquad
     \subfigure[]{\includegraphics[width=5cm]{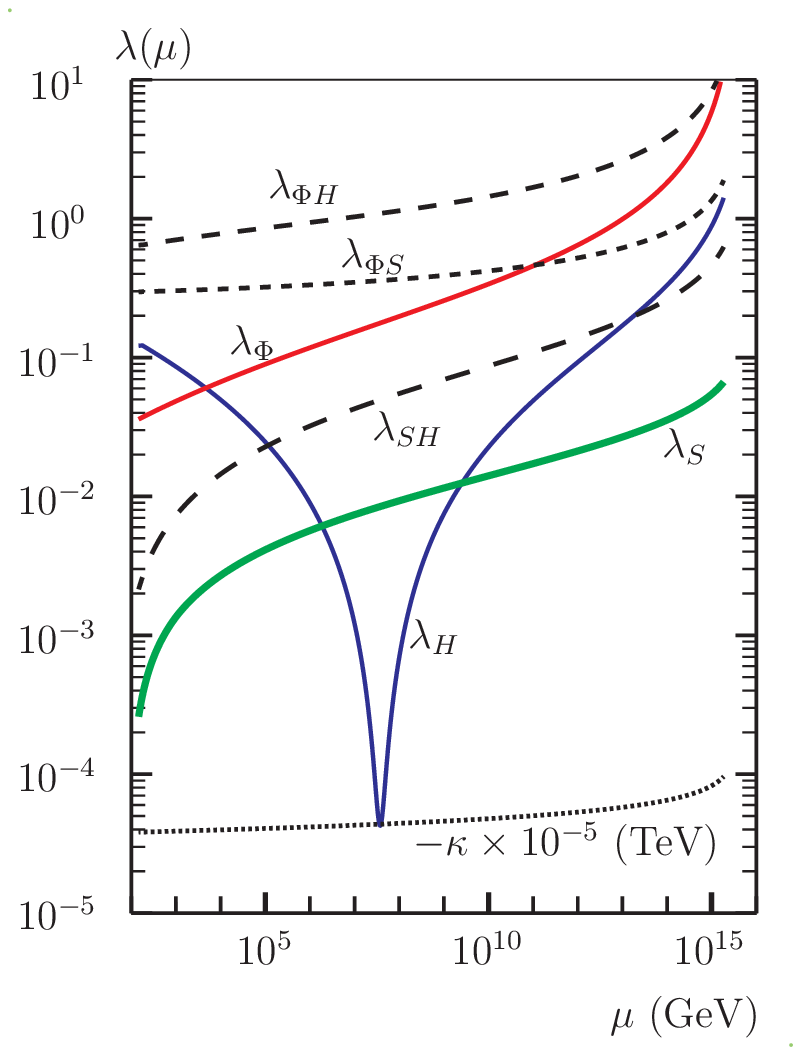}}% \label{fig:RGE4PH}}
       \caption{ Two typical 1-loop RGE running for $\lambda$'s and $\kappa$. (a) Configuration A: $R_{VS}=2.07$ when $\lambda_H$(blue) hits zero.
    (b) Configuration B: $R_{VS}=9.99$  where  $\lambda_\Phi$(red) hits a Landau pole.
 }%
    \label{fig:RGE_example}%
\end{figure}

To demonstrate this we display the details of two typical configurations:
\begin{itemize}
\item Configuration A\\
 $T_{dec}=1.944$GeV, $M_S=27.31$GeV, $\theta=-0.0268$, $M_\rho=2.21$TeV, $\lambda_{SH}=0.000244$,
$\lambda_H=0.12901$, $v_S=6.65$TeV, $\lambda_S=8.55\times 10^{-6}$, $Y_S=0.213$, $\bar{\kappa}=-1.17$TeV, $\lambda_{\phi H}=0.541$, $\lambda_{\phi S}=1.40$, and $\lambda_\phi=0.051$.\\
The scalar sector is stable until $\lambda_H$ becomes negative, and $R_{VS}=2.07$.
Moreover, $\Gamma_S=5.25\times 10^{-6}$ GeV, $Br(S\ra \omega\omega)=0.872$,  $Br(S\ra b\bar{b})=0.108$,
 $Br(S\ra c\bar{c})=0.012$, and $Br(S\ra \tau\bar{\tau})=0.008$.
\item Configuration B\\
$T_{dec}=1.87$GeV, $M_S=67.55$GeV, $\theta=-0.319$, $M_\rho=1.83$TeV, $\lambda_{SH}=0.0011$,
$\lambda_H=0.1201$, $v_S=12.1$TeV, $\lambda_S=1.9\times 10^{-5}$, $Y_S=0.117$, $\bar{\kappa}=-0.23$TeV, $\lambda_{\phi H}=0.641$, $\lambda_{\phi S}=0.296$, and $\lambda_\phi=0.0334$.\\
In this example, $R_{VS} = 9.99$ where $\lambda_\Phi$ hits a Landau pole, $\Gamma_S=2.63\times 10^{-4}$ GeV, $Br(S\ra \omega\omega)=0.072$,  $Br(S\ra b\bar{b})=0.783$, $Br(S\ra c\bar{c})=0.081$, and $Br(S\ra \tau\bar{\tau})=0.052$.

 \end{itemize}

 The RGE running of scalar quartic couplings and $\kappa$ for configuration-A and B  are shown in Fig.\ref{fig:RGE_example}.

\begin{figure}[ht]
    \centering
     \subfigure[]{\includegraphics[width=4cm]{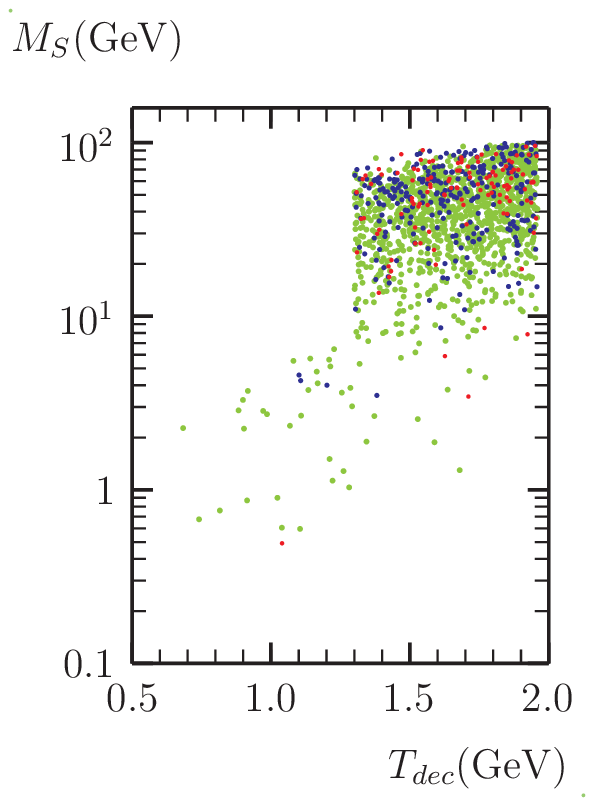}}%
       \qquad
    \subfigure[]{\includegraphics[width=4cm]{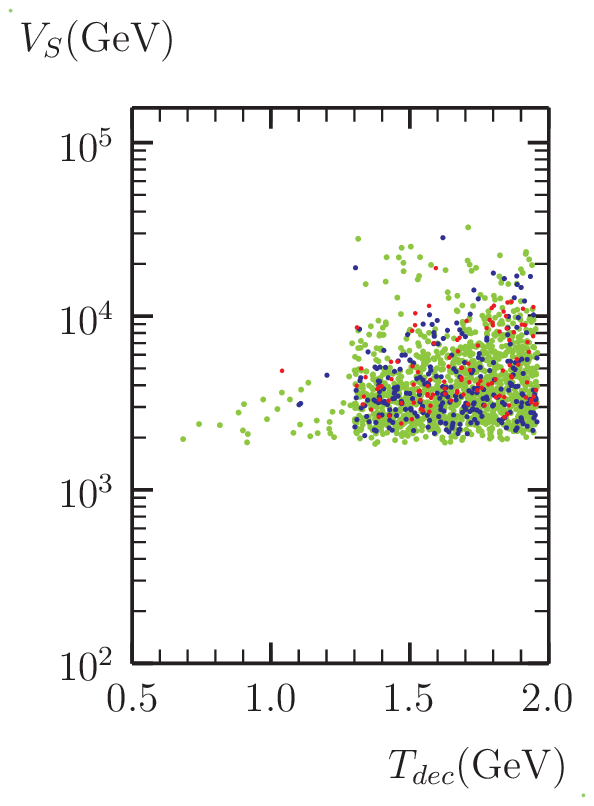}}%
        \qquad
    \subfigure[]{\includegraphics[width=4cm]{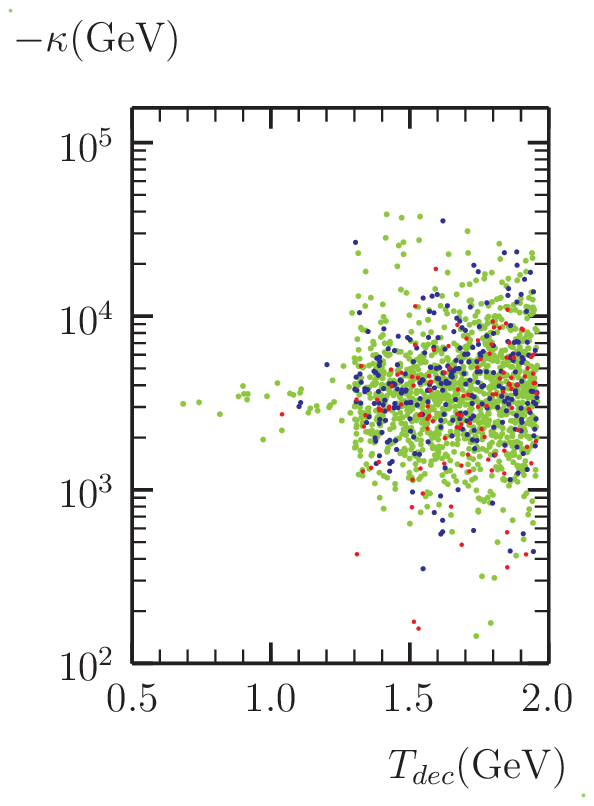}}%
    \qquad
          \subfigure[]{\includegraphics[width=4cm]{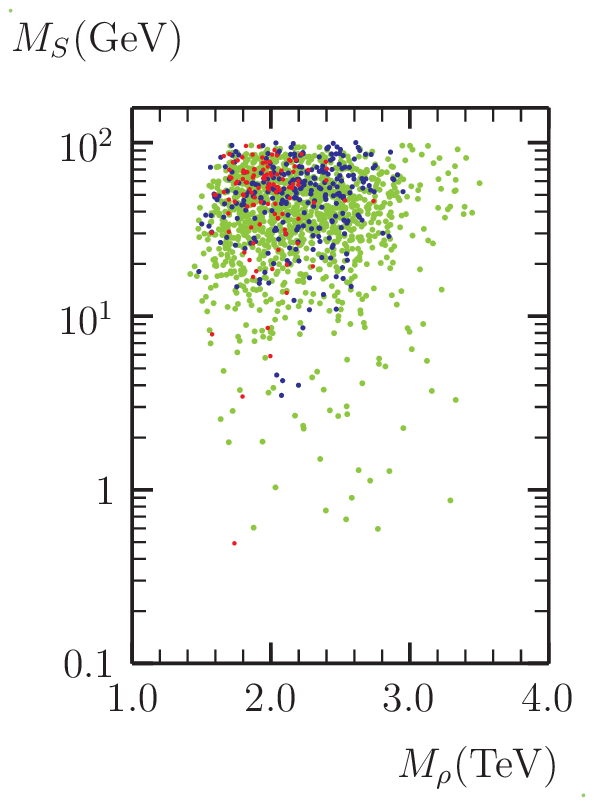}}%
    \qquad
         \subfigure[]{\includegraphics[width=4cm]{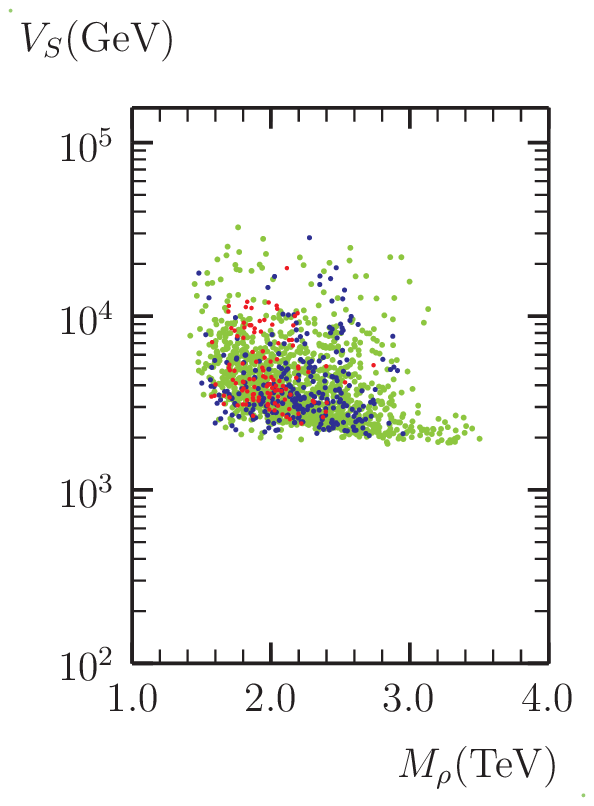}}%
   \qquad
             \subfigure[]{\includegraphics[width=4cm]{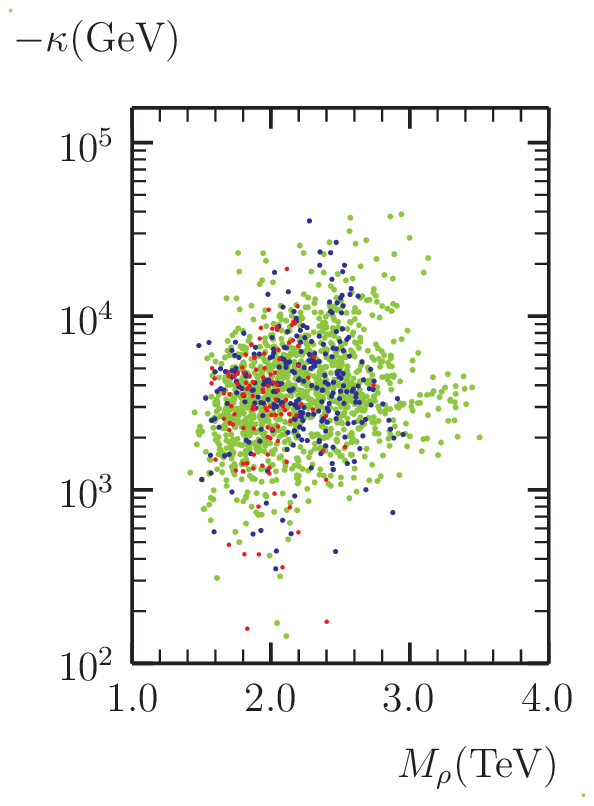}}%
    \caption{Scatter plots for $M_S$, $v_S$, and $-\kappa$ v.s. $T_{dec}$(a-c) and $M_\rho$(d-f).
    Where the color codes are: Green = $2< R_{VS}<4$,  Blue: $4<R_{VS}<6$, and Red: $R_{VS}>6$. }%
    \label{fig:MS_VS_Kappa_listplot}%
\end{figure}

\begin{figure}[ht]
    \centering
          \subfigure[]{\includegraphics[width=4.5cm]{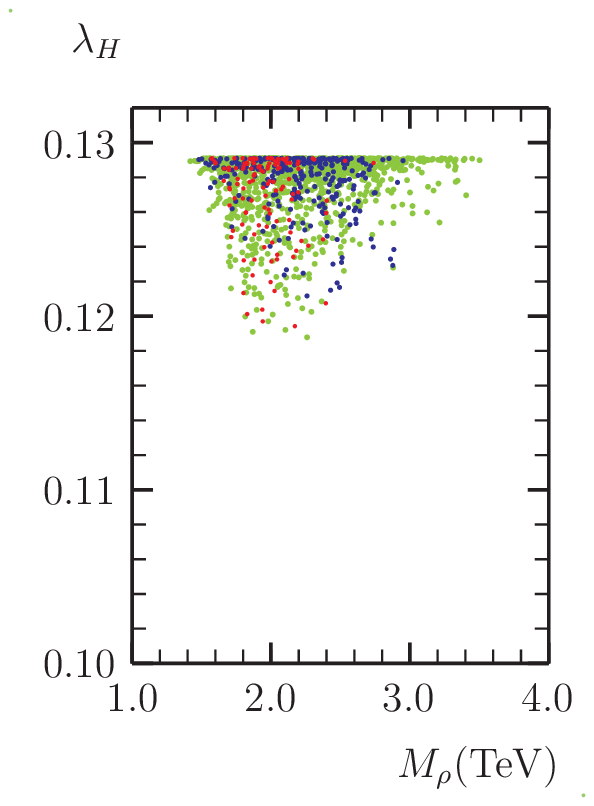}}%
    \qquad
      \subfigure[]{\includegraphics[width=4.5cm]{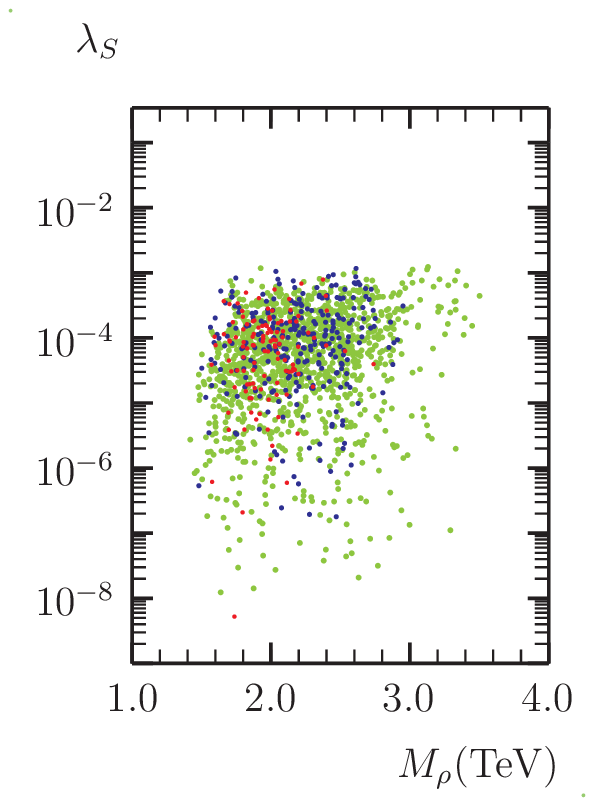}}%
          \qquad
            \subfigure[]{\includegraphics[width=4.5cm]{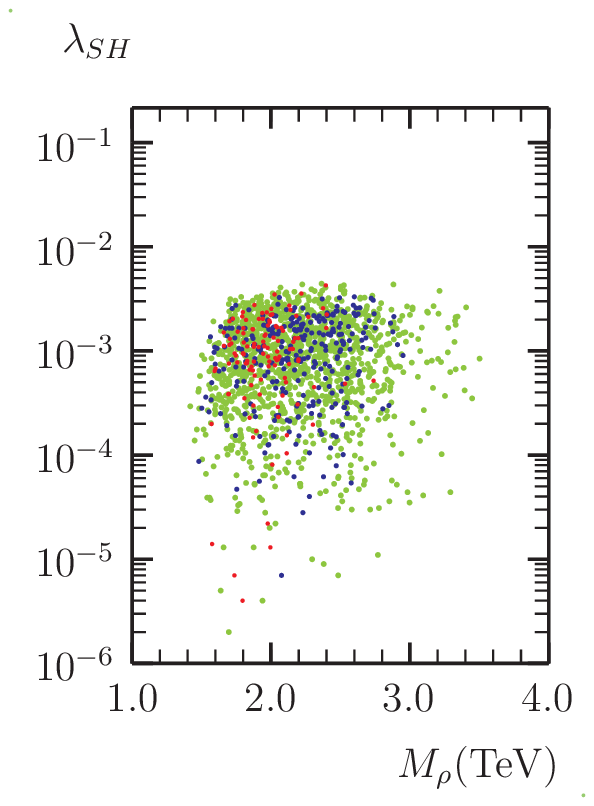}}%
          \qquad
      \subfigure[]{\includegraphics[width=4.5cm]{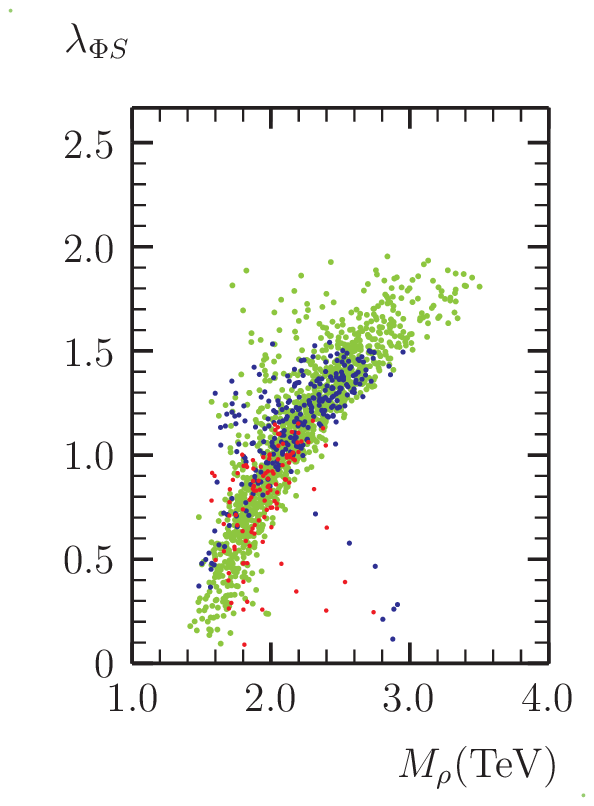}}%
    \qquad
      \subfigure[]{\includegraphics[width=4.5cm]{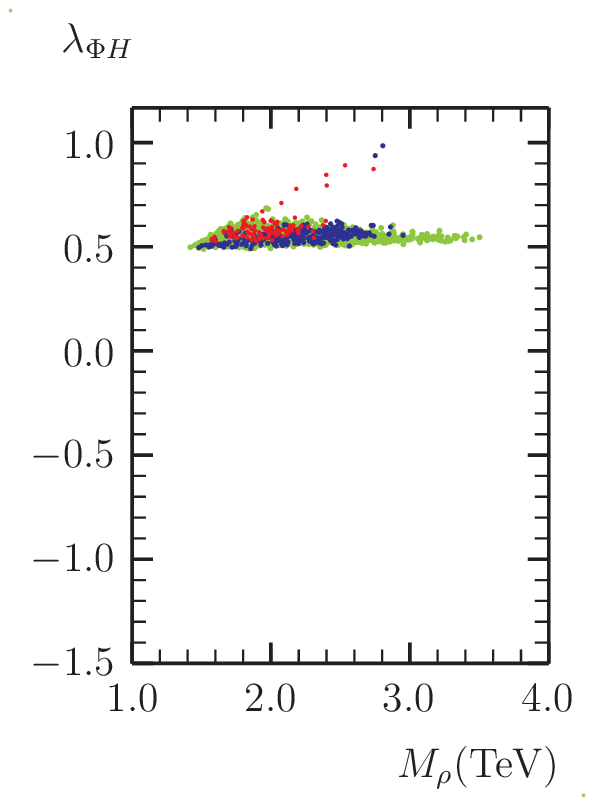}}%
          \qquad
            \subfigure[]{\includegraphics[width=4.5cm]{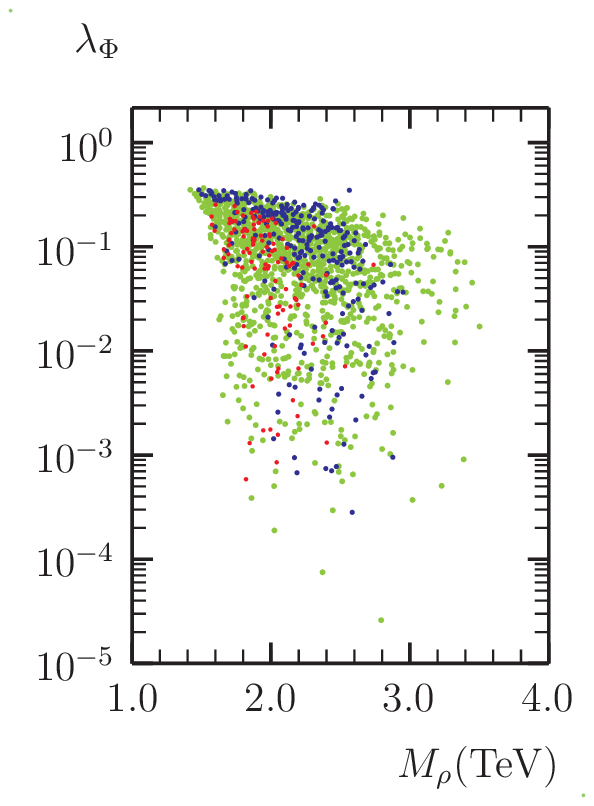}}%
          \qquad
                \subfigure[]{\includegraphics[width=4.5cm]{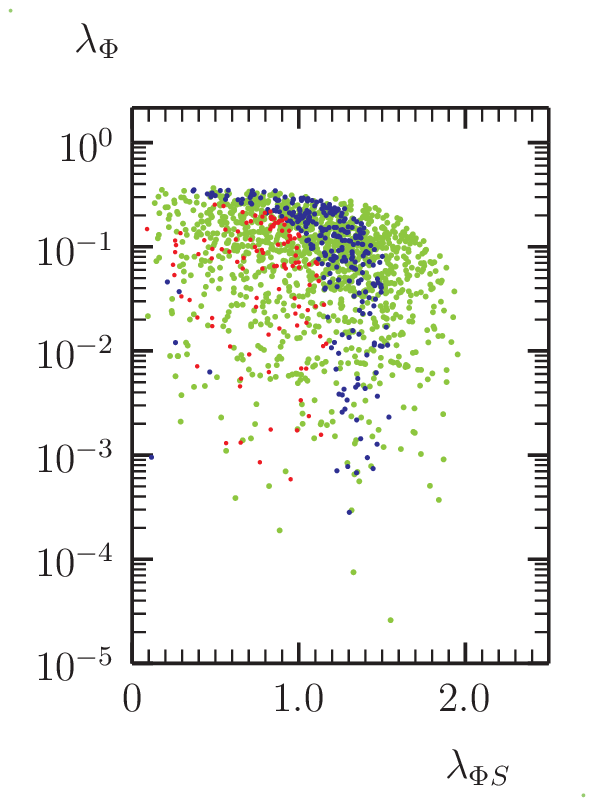}}%
    \qquad
      \subfigure[]{\includegraphics[width=4.5cm]{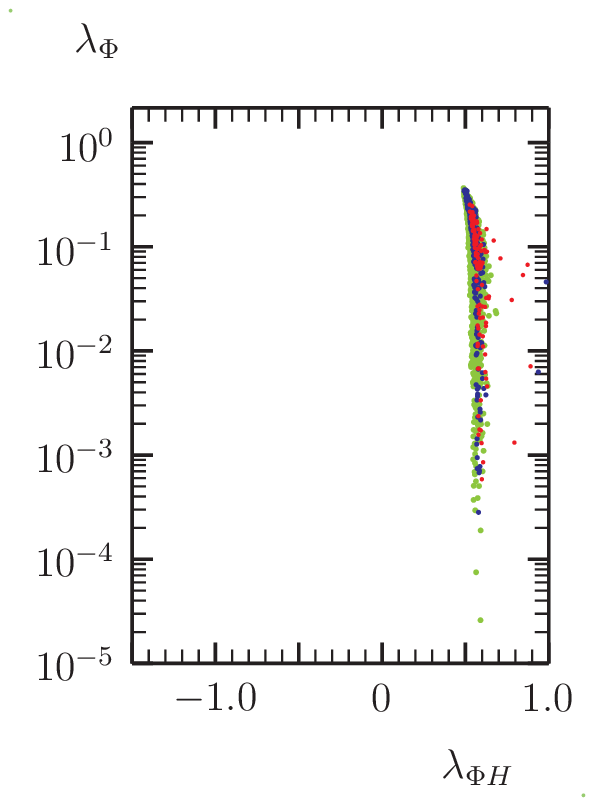}}%
          \qquad
            \subfigure[]{\includegraphics[width=4.5cm]{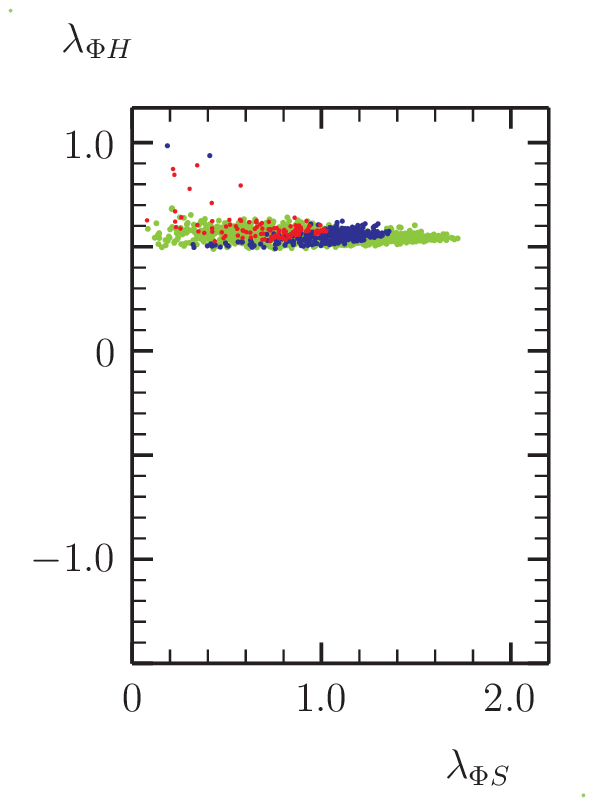}}%

    \caption{Scatter plots for scalar couplings v.s. $M_\rho$.}%
    \label{fig:lambdas_Mrho_listplot}%
\end{figure}

The results are summarized in  Figs.(\ref{fig:MS_VS_Kappa_listplot},\ref{fig:lambdas_Mrho_listplot},\ref{fig:DM_ann_BR_listplot}) where the
green dots represent the configurations with $ 2 <R_{VS}<4$, the blue dots represent the ones with $ 4< R_{VS}<6$, and the red dots show those with $ R_{VS} >6$.
In short, with very mild  fine tuning, $\sim 10^{-2}$, the new scalar degrees of freedom can help to stabilize the SM up to the GUT scale.

The other features of our numerical results can be summarized as follow:
\begin{itemize}
  \item It is easier to find solutions when $T_{dec}\gtrsim 1.3$ GeV and $M_S, V_S, \kappa$ are not very sensitive to $T_{dec}$, Fig.\ref{fig:MS_VS_Kappa_listplot}(a-c). Moreover, $R_{VS}$ does not seem to depend on $T_{dec}$.
      So we will focus instead the parameters dependance on $M_\rho$.
  \item Solutions show that $M_\rho$ is in between roughly $1.5-4$ TeV and center at around $2.5$TeV with larger $R_{VS}$, Fig.\ref{fig:MS_VS_Kappa_listplot}(d-f). Although  the range that $M_\rho \in \{0.5,4\}$TeV is scanned in our numerical study, we found no solutions with $M_\rho \lesssim 1.5$ TeV.
  \item $M_S$ is mainly in the $20-10^2$ GeV range, Fig.\ref{fig:MS_VS_Kappa_listplot}(d).
  \item $V_S$ and $-\kappa$ center at around $2-20$ TeV, Fig.\ref{fig:MS_VS_Kappa_listplot}(d-e).
  \item From Fig.\ref{fig:lambdas_Mrho_listplot}(a-c), we see that $\lambda_H\in \{0.118,0.130\}$ and peaks at around $\sim 0.129$, the SM value; $\lambda_S\in \{10^{-9},10^{-3}\}$ and peaks at around $\sim 10^{-4}$;  $\lambda_{SH}\in \{10^{-6},10^{-2}\}$ and peaks at around $\sim 10^{-3}$. The $M_\rho$-dependance is weak.
  \item  $\lambda_{\Phi S}$ rises quickly from zero when $M_\rho\gtrsim 1.5$TeV, Fig.\ref{fig:lambdas_Mrho_listplot}(d).
  \item $\lambda_{\Phi H}$ peaks at around  $+ 0.5$  and it is not very sensitive to $M_\rho$, $\lambda_{\Phi}$, or $\lambda_{\Phi S}$, Figs.\ref{fig:lambdas_Mrho_listplot}(e,h,i).

  \item $\lambda_\Phi\sim {\cal O}(0.1)$ and depends on $M_\rho$ weakly,  Fig.\ref{fig:lambdas_Mrho_listplot}(f).
  \item The upper bound of $\lambda_\Phi$ depends on $\lambda_{\Phi S}$ near $\lambda_{\Phi S}\gtrsim 1.0$,  Fig.\ref{fig:lambdas_Mrho_listplot}(g).
\end{itemize}

\subsection{RGE running}
The coupled RGEs are highly entangled and it is not easy to have an insight by cursory inspections.
 Now with the help of numerical results, we can gain some qualitative understandings the behaviors the solutions to these RGEs.
At  1-loop the beta function for $\lambda_\Phi$ yields values that are always positive in the interested energy range ( see Eq.(\ref{eq:RGE}b)), so we only need to worry about the vacuum instabilities of $\lambda_H$  and $\lambda_S$.
The SM part of the 1-loop beta function for $\lambda_H$, i.e. the righthand side of Eq.(\ref{eq:RGE}a) except the last two terms, takes a value $\simeq -2.1$ at around ${10^2}$ GeV. Since the Majoron decouples at around a few GeV or less, $\lambda_{SH}^2 <10^{-3}$ even for $ M_s\sim 100$ GeV, see Eq.(\ref{eq:dec}). Therefore,  $\lambda_{\Phi H}$ plays the leading role of  improving $\lambda_H$ stability.
By linear extrapolation,  the beta function
needs an extra  $\sim 2.1- (16 \pi^2)\times \lambda_H/ [ \ln (10^2 \mu_{VS}^{SM})^2 - \ln M_Z^2 ] \sim +1.2$ contribution from the $\lambda_{\Phi H}$-term to move up the Higgs scalar potential stability limit at  $\mu_{VS}^{SM}$ to $100\times \mu_{VS}^{SM}$ That amounts to $\lambda_{\Phi H} \sim \pm \sqrt{2.4} \sim {\cal O}( \pm 1)$. However, the negative solution is not viable. Because the sizable negative $\lambda_{\Phi H}$ will quickly drive the $\lambda_{HS}$ into large negative value during the RGE running such that  $\lambda_{HS} <-2\sqrt{\lambda_H \lambda_S}$ and violates a vacuum stability condition.  This estimate agrees with the feature we found in the numerical study  that $\lambda_{\Phi H}$ centers at around $ + 0.5$ and it is not very sensitive to other parameters.

\begin{figure}[ht]
    \centering
   \subfigure[]{\includegraphics[width=5cm]{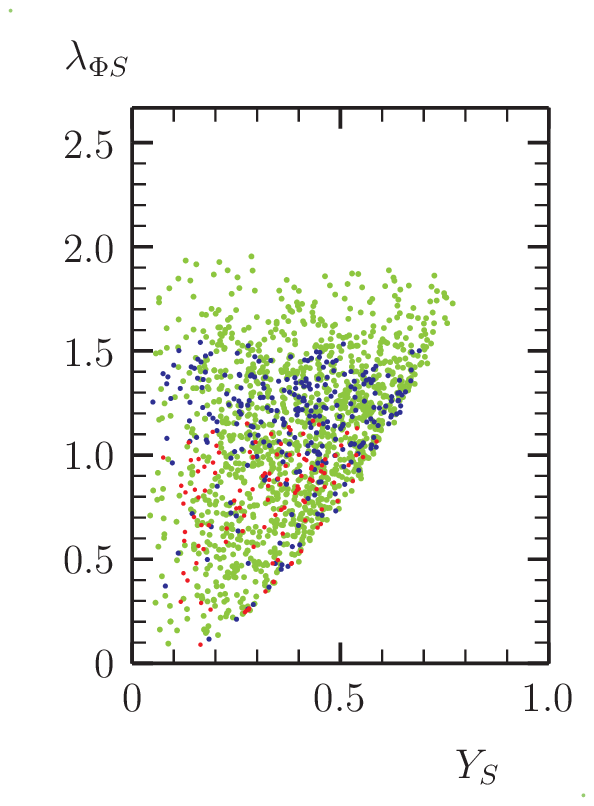}}%
    \qquad
     \subfigure[]{\includegraphics[width=5cm]{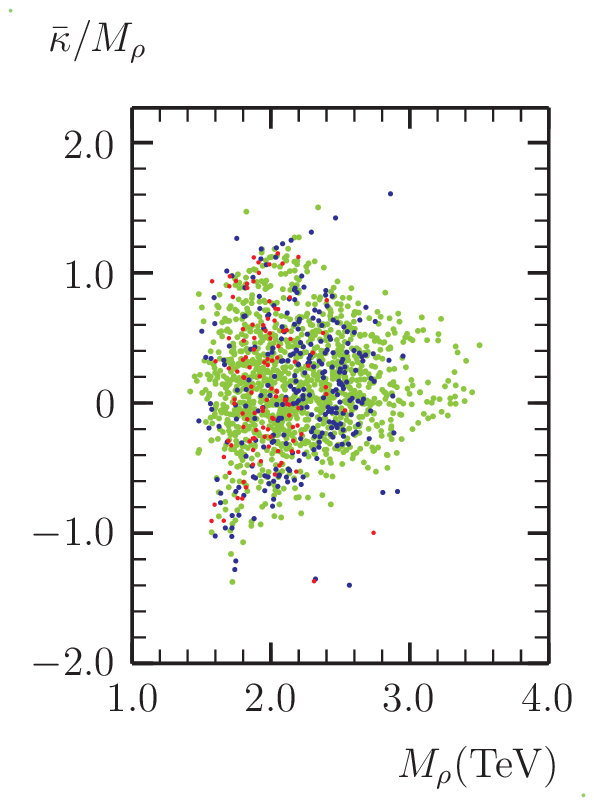}}

    \caption{(a) Correlation between $\lambda_{\Phi S}$ and the Yukawa coupling $Y_S$. (b)  $\bar{\kappa}/M_\rho$  v.s. $M_\rho$.}%
    \label{fig:lam_Ys}%
\end{figure}

Now, the lepton number breaking scale $v_S$ is pushed up by the small $\lambda_{SH}$, Eq.(\ref{eq:VS_mass_basis}), and the
$\lambda_S$ is brought down by increasing $v_S$, Eq.(\ref{eq:lam_S_mass_basis}). Therefore,  $\lambda_{\Phi S}$ governs the stability of $\lambda_S$  and  it competes with the negative contribution from $Y_S$ in the beta function for  $\lambda_S$, Eq.(\ref{eq:RGE}c).  Roughly speaking,  one needs  $\lambda_S^2 \gtrsim 16 Y_S^4$
to stabilize the $\lambda_S$-vacuum.  Moreover, the positivity requirement that $ \lambda_{\Phi S}>-4\sqrt{\pi \lambda_S}$ eliminates the negative solution for $\lambda_{\Phi S}$. This
boundary $ \lambda_{\Phi S} \gtrsim 4 Y_S^2$ can be clearly seen in our numerical result, Fig.\ref{fig:lam_Ys}(a).

In addition to the issue of vacuum stability, the scalar sector in our model also introduces the
Landau pole problem. As already mentioned, the beta function of $\lambda_\Phi$ is always positive and it could leads to Landau pole.
Assuming that $\lambda_{\Phi H}$ and $\lambda_{\Phi S}$ are more or less constant during the RGE running between $t_0$ and $t$, then Eq.(\ref{eq:RGE}b) admits an exact solution for $\lambda_\Phi$:
\beq
\lambda_\Phi(t)= \sqrt{\frac{\alpha_2}{\alpha_1}}\tan[\sqrt{\alpha_1 \alpha_2} t+ \alpha_3]\,,
\eeq
where $\alpha_1=10/(16\pi^2)$, $\alpha_2=(\lambda^2_{\Phi H}+\lambda^2_{\Phi S}/2 )/(16\pi^2)$, and $\alpha_3=\tan^{-1}(\lambda_\Phi(t_0)\sqrt{\alpha_1/\alpha_2})$. The Landau pole appears at $t_{Landau}$ when the argument inside tangent
becomes $\pi/2$. Or,
\beq
t_{Landau}= {16 \pi^2 \over \sqrt{10\lambda_{\Phi H}^2 + 5\lambda_{\Phi S}^2  }}
\left[ \frac{\pi}{2}-\tan^{-1} \left( \lambda_\Phi(t_0)\sqrt{{10 \over \lambda_{\Phi H}^2 + \lambda_{\Phi S}^2/2 }}  \right) \right]\,.
\eeq
From this expression, it is clear that small $\lambda_\Phi$ and $\lambda_{\Phi S}$ are preferred if one wishes to have a large $R_{VS}$ before hitting a Landau pole or even a Landau pole beyond $M_{Pl}$.
This agrees very well with what we have observed in the numerical experiment, see Fig.\ref{fig:lambdas_Mrho_listplot}(g) and
Fig.\ref{fig:lam_Ys}(a).
We did two simple numerical checks with the Configuration-B by modifying: (1) $\lambda_\Phi \Rightarrow \lambda_\Phi+0.01$,
or (2) $\lambda_{\Phi S} \Rightarrow \lambda_{\Phi S}+0.01$.
Originally, a Landau pole happens at $1.85\times 10^{15}$ GeV. But now  the slight modification makes
$\lambda_\Phi$ blows up at $1.204\times 10^{15}$ GeV and  $1.78\times 10^{15}$ GeV
respectively.

\subsection{A second look at DM annihilation}
\begin{figure}[ht]
    \centering
 \subfigure[]{\includegraphics[width=4.5cm]{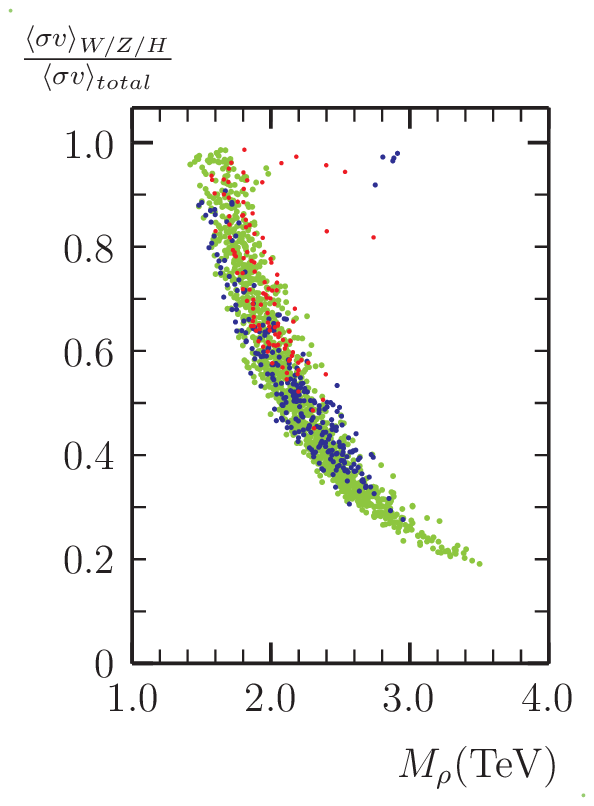}}%
    \qquad
     \subfigure[]{\includegraphics[width=4.5cm]{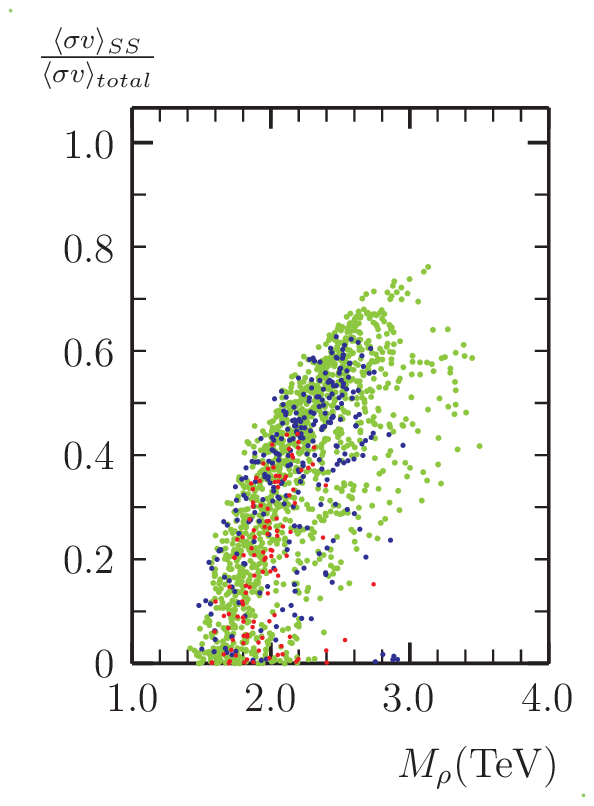}}
  \qquad
     \subfigure[]{\includegraphics[width=4.5cm]{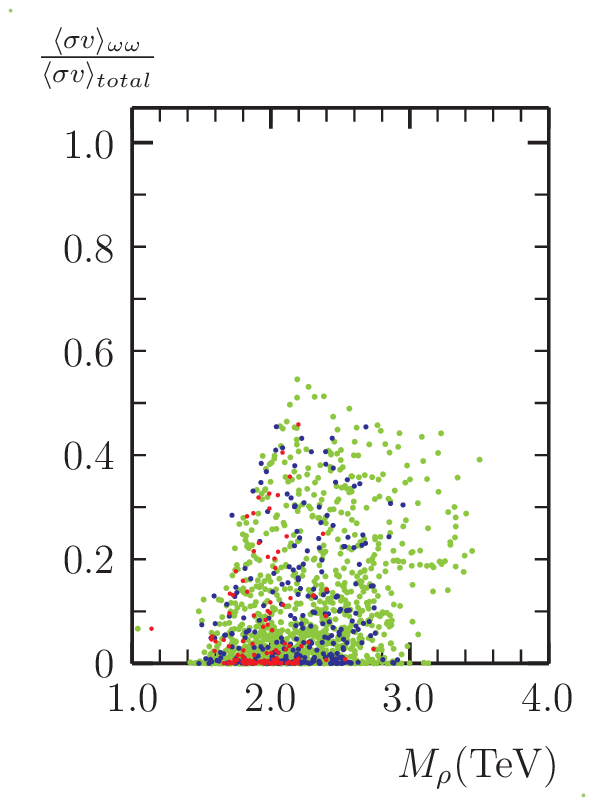}}

    \caption{ The fractions of DM annihilation into (a) the SM final states, (b) $SS$ pair, and (c) $\omega\omega$  v.s. $M_\rho$.}%
    \label{fig:DM_ann_BR_listplot}%
\end{figure}

Given that  $ \lambda_S, \lambda_{SH}\ll 1$ and $M_\rho\gg M_W, M_Z, M_H$, the DM annihilation into SM final states cross sections are mainly controlled by $\lambda_{\Phi H}$,
see Eq.(\ref{eq:sigv},\ref{eq:coupling_mass_basis}).
Furthermore, due to the small mass ratio $x_f=m_f/M_\rho$, the $\rho\rho\ra \bar{f} f$ takes up only a tiny fraction of the total $\langle \sigma v\rangle_{total}= 2.5\times 10^{-9}(GeV)^{-2}$ , i.e. between $0.2\% \sim 0.8\%$ overall, and peaks at around $0.3\%$ when $M_\rho \sim 2.5$ TeV.
 For $|\lambda_{\Phi H}|\sim 0.5$, the total annihilation cross section of $\rho\rho\ra W^+W^-, ZZ, HH$
can be estimated  to be
\beqa
\langle \sigma v\rangle_{W/Z/H} \equiv \langle \sigma v\rangle_{W^+W^-}+\langle \sigma v\rangle_{ZZ}+\langle \sigma v\rangle_{HH}
\sim \frac{1}{64\pi}\frac{\lambda_{\Phi H}^2}{M_\rho^2}\times[2+1+1]\nonr\\
 \sim 5\times 10^{-9} (GeV)^{-2}\left({\lambda_{\Phi H} \over 0.5}\right)^2 \left({1 \mbox{TeV} \over M_\rho}\right)^2\,.
\eeqa
And it is clear now why this model prefers a heavy DM ( $\gtrsim 1.4$ TeV ) after taking into account the RGE running and the issue of vacuum stability. When DM is relatively light, close to $1.4$ TeV, the total cross section is saturated by the channels with SM final states. Since  $|\lambda_{\Phi H}|$ is not sensitive to $M_\rho$, we immediately expect that $\langle \sigma v\rangle_{W/Z/H}/ \langle \sigma v\rangle_{total}$ is inversely proportional to DM mass squared, as it is shown in Fig.\ref{fig:DM_ann_BR_listplot}(a).
On the other hand, the cross sections of DM annihilation into $\omega$ and $S$ are mainly governed by $\lambda_{\Phi S}$ and $\kappa$.
Only when $M_\rho\gtrsim 2$ TeV the other channels, $\rho\rho\ra SS$ and $\rho\rho\ra\omega\omega$, can make important
contributions to $\langle \sigma v\rangle_{total}$,  see Figs.\ref{fig:DM_ann_BR_listplot}(b,c).

\section{Phenomenology}
\subsection{Extra light scalar $S$ and its mixing with the Higgs boson}
\begin{figure}[ht]
    \centering
 \subfigure[]{\includegraphics[width=4.4cm]{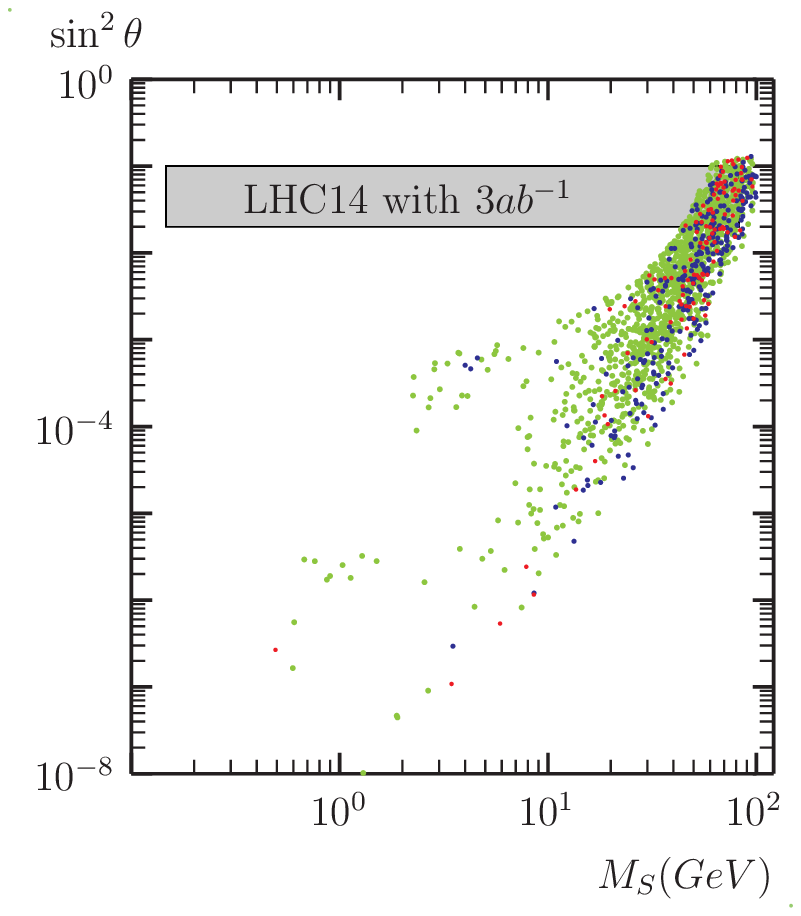}}%
    \subfigure[]{\includegraphics[width=4.4cm]{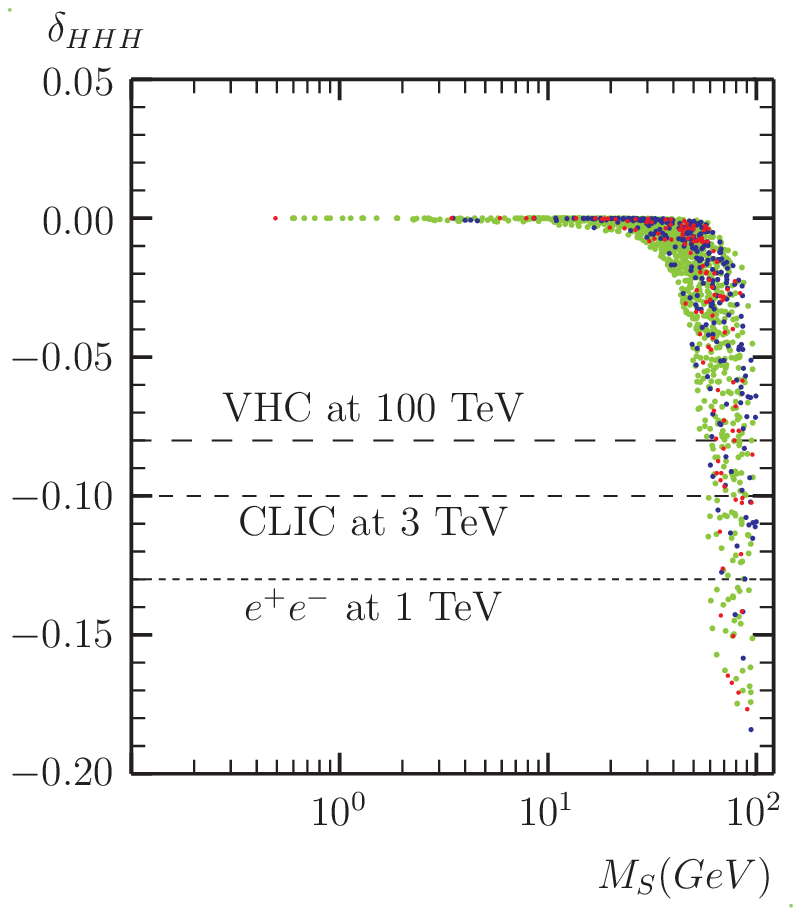}}%
     \subfigure[]{\includegraphics[width=4.4cm]{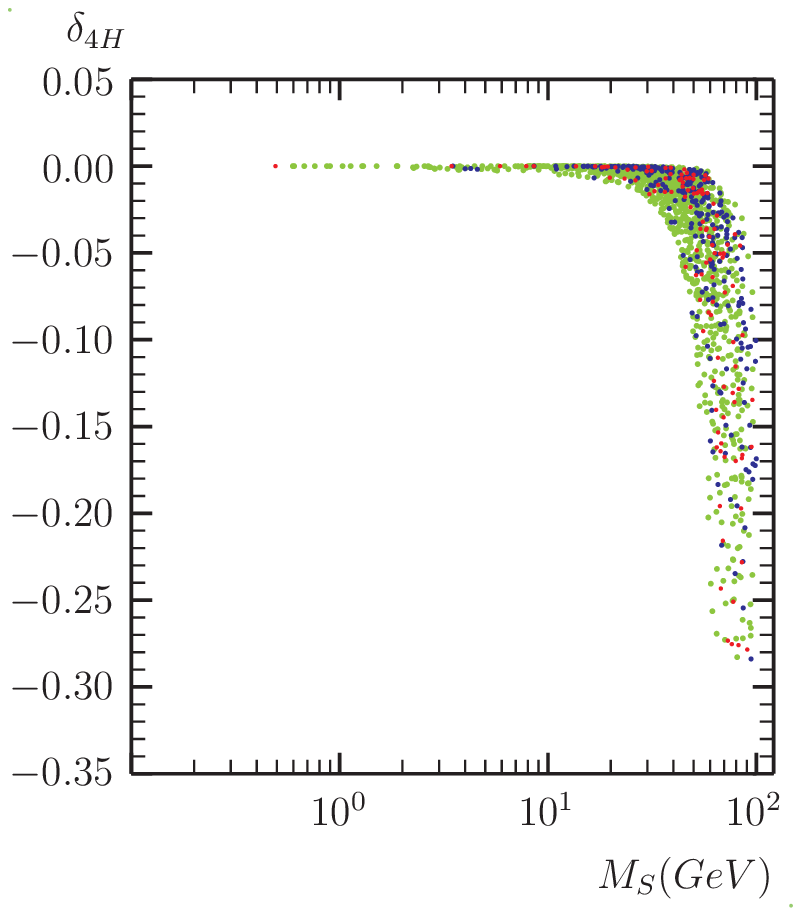}}%
 \caption{(a) Scattering plot for $\sin^2\theta$ v.s. $M_S$. The gray band is the expected sensitivity  by improving the signal strengthes measurement at the LHC14.
 (b)$\delta_{HHH}$, the deviation fraction of the Higgs triple coupling  from the SM value.
 The dash lines represent the expected precisions at the future facilities.
 (c) The deviation of quartic-Higgs coupling in this model.}
 \label{fig:HS_mixing}
\end{figure}

As mentioned in Section II, all signal strengths take a  universal value of $\cos^2\theta$  in our model due to the $H-S$ mixing. In Fig.\ref{fig:HS_mixing}(a), the correlation between $\sin^2_\theta$ and $M_S$ from our numerical study is displayed as well the expected sensitivity of $\sin^2\theta$ by improving the signal strength measurements at the LHC14 with $ 3 ab^{-1}$ luminosity.
The parameter space with $M_S\gtrsim 40$ GeV or equivalently the large mixing angle in our model will be covered by LHC14.
If no detectable deviation  is found, this part of parameter space will be discarded.
On the other hand, if this large mixing region is not excluded by LHC14, the same parameter space can be further directly probed by future facilities as we discuss next.

One immediate consequence of the existence of a neutral scalar of mass few tens of GeV and sizable mixing is that the triple SM Higgs coupling, $\lambda^{SM}_{HHH} = 6\lambda_H v_H=3M_H^2/v_H$,
will be reduced.
The tree level triple Higgs coupling is given in Eq.(\ref{eq:coupling_mass_basis}) as $\lambda_{HHH}$.
In Fig.\ref{fig:HS_mixing}(b), the deviation $\delta_{HHH} = (\lambda_{HHH}-\lambda^{SM}_{HHH})/\lambda_{HHH}^{SM}$ is displayed.
The deviation can be as large as $20\%$ when $M_S \sim 100$ GeV and center around few percents for configurations with better RGE improvement. This Higgs triple coupling  is expected to be probed  to $50\%$  at LHC14 with $3 ab^{-1}$ luminosity\cite{Dawson:2013bba} and $\sim 10\%$ at CEPC\cite{Gomez-Ceballos:2013zzn}.
So  some of the parameter space with $M_S >40$ GeV in this model can be probed at future colliders.

Similarly, the quartic coupling of the SM Higgs will be modified from $\lambda_{4H}^{SM}= 6\lambda_H=3 (M_H/v_H)^2$
to $\lambda_{4H}=6 (\lambda_H c_\theta^4+\lambda_S s_\theta^4)$ in this model. The deviation $\delta_{4H} =(\lambda_{4H}-\lambda_{4H}^{SM})/\lambda_{4H}^{SM}$ could reach $\sim 30\%$ when $M_S \sim 100$ GeV, see Fig.\ref{fig:HS_mixing}(c). Note that for small $\lambda_S$, $\delta_{HHH}\sim (c_\theta^3-1)$ and $\delta_{4H}\sim (c_\theta^4-1)$, therefore $|\delta_{4H}|>|\delta_{HHH}|$.
In principle the quartic coupling could be probed through the triple Higgs production but this cross section is hopelessly small to be searched for at any foreseen future facility.

\subsection{Decays of $S$}

Since now that the mass of $S$ is in the range of few tens to one hundred GeV, more decay channels are opened up than in the case that $M_S< 1$ GeV which was discussed in \cite{CN}.
The various decay widths can be easily derived from the well known ones in the SM.
The widths of the dominant decay channels are given at below:
\beqa
\Gamma_{S\ra ff} &=& \frac{s_\theta^2 N_C^f M_S }{8\pi}\left(\frac{m_f}{v_H}\right)^2 \left(1-4\frac{m_f^2}{M_S^2}\right)^{3/2}\,,\,\,
\Gamma_{S\ra \omega\omega} = \frac{c_\theta^2 M_S^3}{32\pi v_S^2}\,,
\eeqa
where $S\ra b\bar{b}$ and $S\ra \omega\omega $  take up about 90\% of the total decay width for $M_S <M_W, M_Z$.
There are also $S\ra gg$ and $S\ra \gamma\gamma$ decays  induced at the 1-loop level:
\beqa
\Gamma_{S\ra gg} &\sim & \frac{s_\theta^2 \alpha_S^2 M_S}{72 \pi^3} \left(\frac{M_S}{v_H}\right)^2 \,,\,\,
\Gamma_{S\ra \gamma\gamma} \sim \frac{s_\theta^2 \alpha^2 M_S}{16 \pi^3} \left(\frac{M_S}{v_H}\right)^2\,,
\eeqa
where we only keep the 1-loop top quark contribution for $S\ra gg$ decay.
For $S\ra \gamma\gamma$, the $W-$loop contribution dominates over the top-loop contribution and the two have opposite sign which give rise to $O(1)$ loop factor which we neglect
in both cases. However, the resulting branching ratio is smaller than $10^{-2} (10^{-4})$ for $S\ra gg(\gamma\gamma)$  and can be ignored.
When $M_S> M_W, M_Z$, the 3-body decays $S\ra W W^*\ra W f \bar{f}'$ and $S\ra Z Z^*\ra Z f \bar{f}$ are opened up and start to play a role.
The total decay widths are
\beqa
\Gamma_{S\ra W^\pm ff'} &\sim & \frac{ 3 s_\theta^2 M_S}{32\pi^3}\left(\frac{M_W}{v_H}\right)^4 F\left(\frac{M_W}{M_S}\right)\,,\\
\Gamma_{S\ra Z ff}  &\sim & \frac{  s_\theta^2 M_S}{128\pi^3}\left(\frac{M_Z}{v_H}\right)^4
\left[6-12s_W^2+\frac{152}{9}s_W^4\right] F\left(\frac{M_Z}{M_S}\right)\,,
\eeqa
where
\beqa
F(x)=-|1-x^2|\left( \frac{47}{2}x^2-\frac{13}{2}+\frac{1}{x^2}\right)
-3(1-6x^2+4x^4)|\ln x|\nonr\\
+3{1-8x^2+20x^4 \over \sqrt{4x^2-1}}\cos^{-1}\left(\frac{3x^2-1}{2x^3}\right)\,.
\eeqa

We have summed over all light final states and treat them as massless particles and
and we excluded the $S\ra W t b $, $S\ra Z b\bar{b}$, and $S\ra Z t\bar{t}$ modes since they
are either kinematically forbidden or suppressed.
Above the mass thresholds, the branching ratio for $S\ra WW^* (ZZ^*)$ can reach $\sim 10^{-5}(10^{-6})$ when $M_S\sim 100$ GeV and can be completely ignored.

\begin{figure}[ht]
    \centering

    \subfigure[]{\includegraphics[width=4cm]{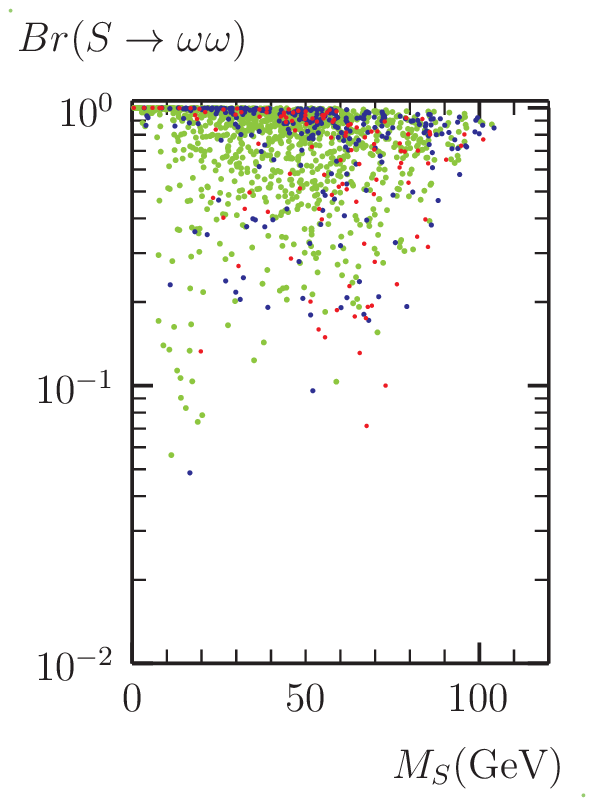}}%
    \qquad
     \subfigure[]{\includegraphics[width=4cm]{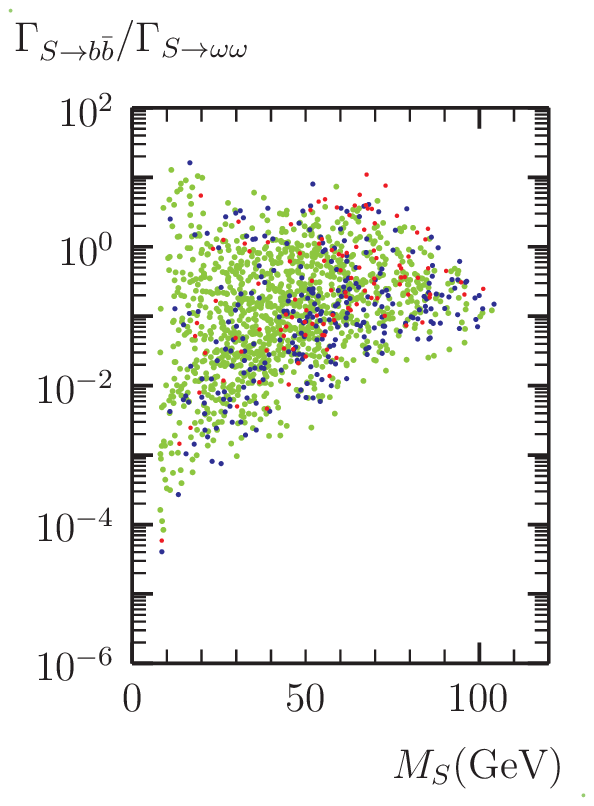}}%
    \qquad
       \subfigure[]{\includegraphics[width=4cm]{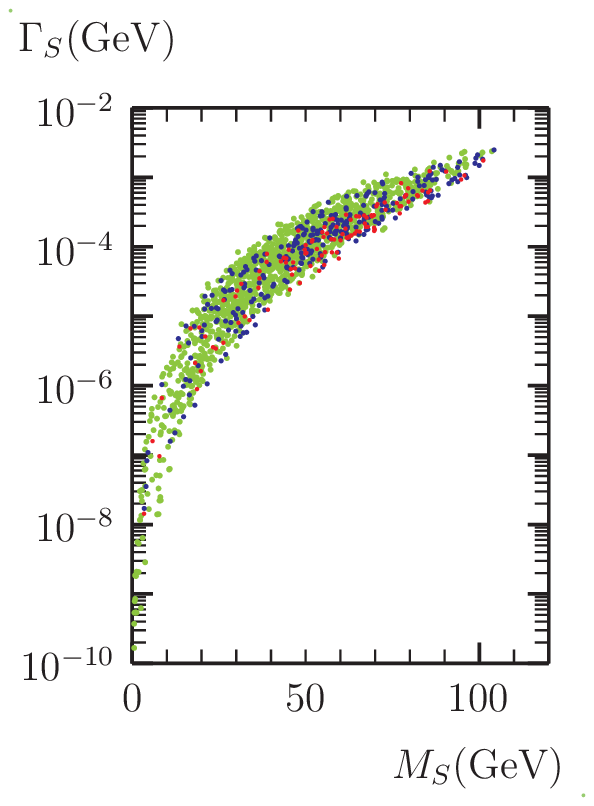}}
    \caption{ (a)Branching ratio of $S\ra \omega\omega$.  (b) Ratio of $\Gamma_{S\ra b\bar{b}}/\Gamma_{S\ra\omega\omega}$. (c) Total decay width of $S$.  }
    \label{fig:S_decay}
\end{figure}

Note that all the decay widths can be fully determined by a given set of $\theta, v_S$ and $M_S$.
In all, the branching ratios and the total decay width are displayed in Fig.\ref{fig:S_decay}.
One can see that in most of the parameter space, $S\ra \omega\omega$ is the dominate decay channel
which has the invisible final states. Even for those configurations with sizable $S\ra b\bar{b}$  decay
branching ratios still suffer from the small production cross section and  make the study of $S$ a challenging task.

\subsection{Using rare $Z,W$ decays to probe light scalar $S$ }
\begin{figure}[ht]
    \centering
   \subfigure[]{\includegraphics[width=4cm]{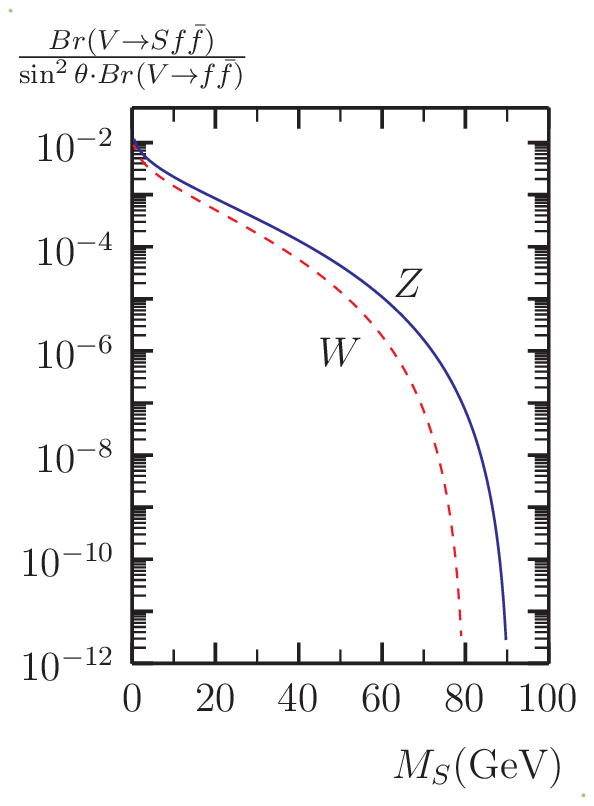}}%
    \qquad
     \subfigure[]{\includegraphics[width=4cm]{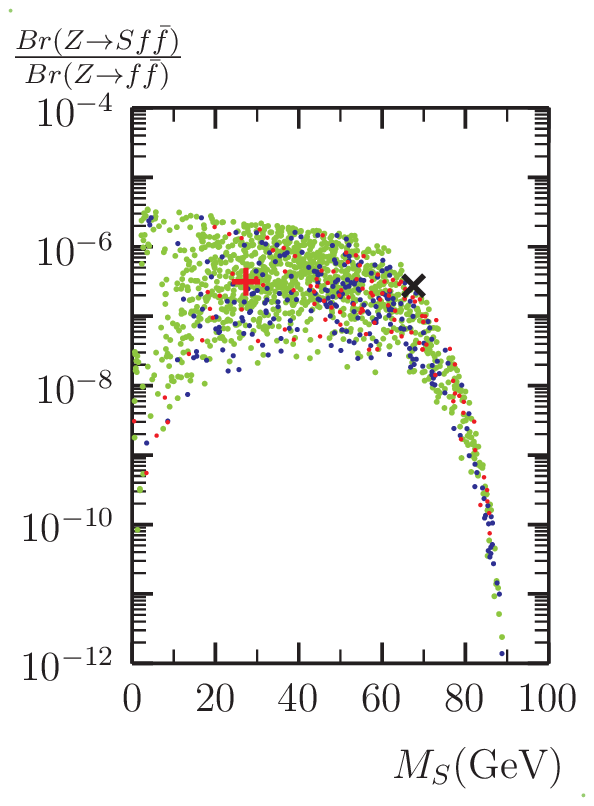}}%
    \qquad
       \subfigure[]{\includegraphics[width=4cm]{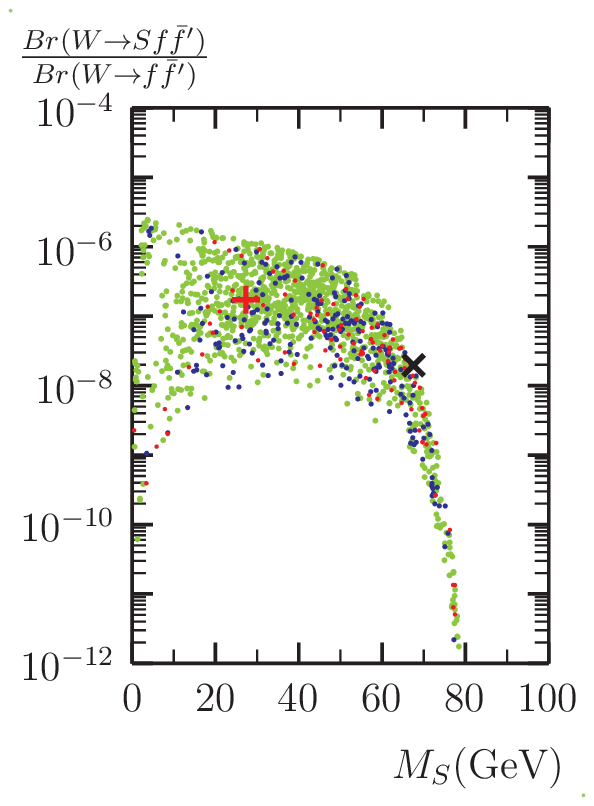}}
    \subfigure[]{\includegraphics[width=4cm]{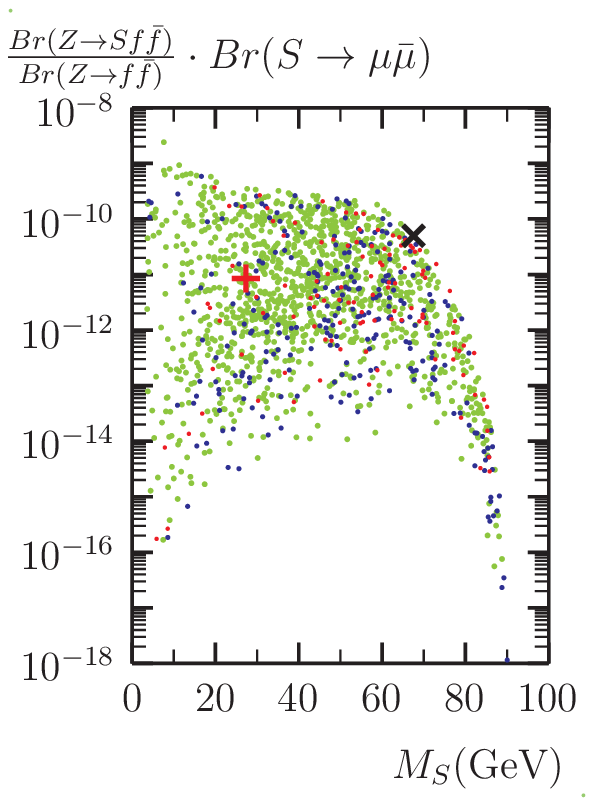}}%
    \qquad
     \subfigure[]{\includegraphics[width=4cm]{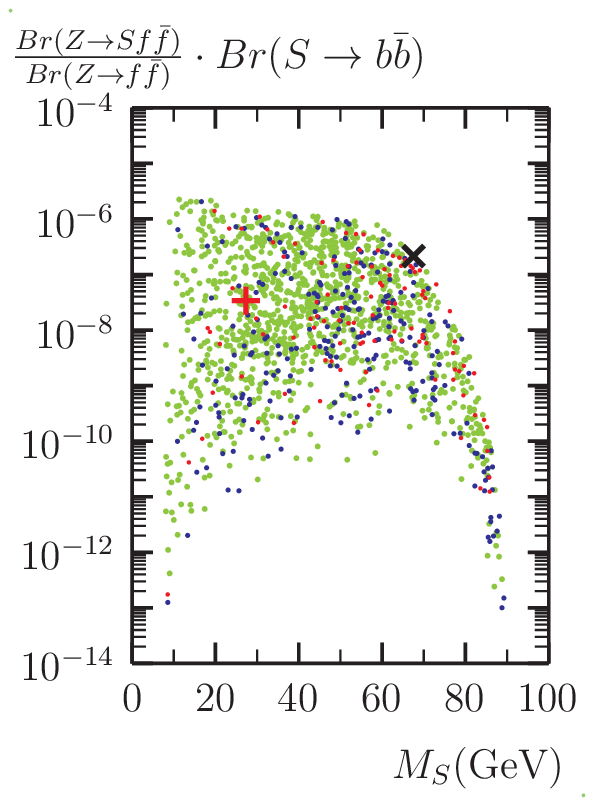}}%
    \qquad
       \subfigure[]{\includegraphics[width=4cm]{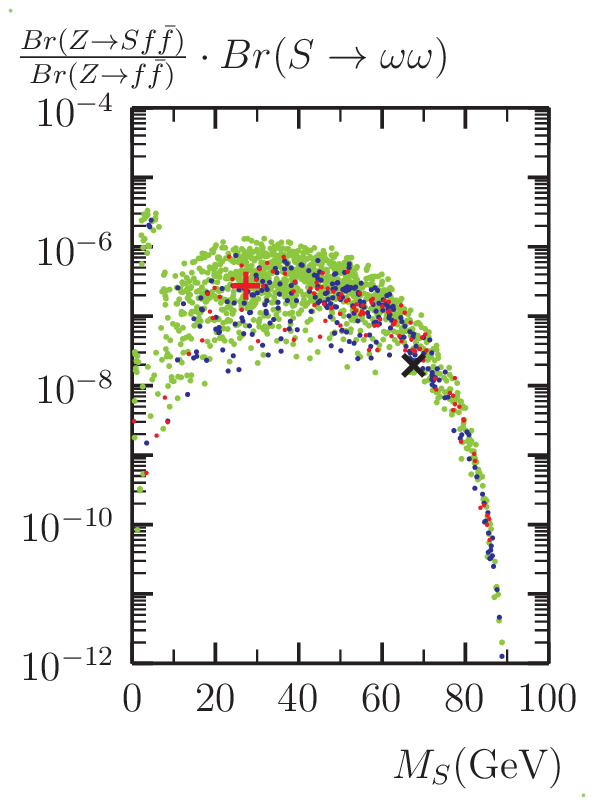}}
    \caption{ (a)Branching ratio modulated the mixing v.s. $M_S$.  (b) Branching ratio for $Z\ra S f\bar{f}$. (c) Branching ratio for $W\ra S f\bar{f}'$. (d,e,f) The branching ratios with $S$ decays into $\mu\bar{\mu}, b\bar{b}$ and $\omega\omega$ pair v.s. $M_S$ from the numerical scan. The corresponding locations for configuration-A and -B are indicated by the red daggers and black crosses respectively. }
    \label{fig:ZSff_decay}
\end{figure}

If $M_S$ below the mass of the Z or W  bosons we can use their rare decays to probe its existence.
This is buoyed  by the expected production of $10^{12-13}$ Z bosons and $10^{8} W^+W^-$ pairs
 per year at the Future Circular Collider involving $e^+e^-$ collisions (FCC-ee)(at $\sqrt{s}=90,160$ GeV with multi-$ab^{-1}$ luminosity \cite{Gomez-Ceballos:2013zzn}. These machines will allow measurements of the properties of the SM gauge bosons at
 unprecedented precision.
Since the SM gauge bosons couple to $S$ though the $S-H$ mixing, the $Z\ra f\bar{f} S$
and $W\ra f\bar{f}' S$ decays are now opened. These decays branching ratios are given by \cite{HiggsHunter}
\beqa
{Br(Z\ra S f\bar{f})\over Br(Z\ra f\bar{f}) }= \frac{g^2 \sin^2\theta}{192\pi^2\cos^2\theta_W}
\left[ {3 r_Z (r_Z^4-8r_Z^2+20)\over \sqrt{4-r_Z^2}}\cos^{-1}\left( \frac{r_Z(3-r_Z^2)}{2}\right)\right.\nonr\\
\left.-3(r_Z^4-6r_Z^2+4)\ln r_Z-\frac{1}{2}(1-r_Z^2)(2r_Z^4-13r_Z^2+47) \right]
\eeqa
for $Z\ra f\bar{f} S$ where $r_Z= M_S/M_Z$, and a similar expression for $Br(W\ra S f\bar{f}')\over Br(W\ra f\bar{f}')$
by substituting $g^2/\cos^2\theta_W$ with $g^2$ also $r_Z\ra r_W=M_S/M_W$. Clearly, the mixing $\sin^2\theta$ plays a pivot role to determine the size of branching ratios. In Fig.\ref{fig:ZSff_decay}(a), we display the $V\ra S f{f}$ branching ratio normalized by $V\ra f\bar{f}$ modulated with the mixing.
And the outcome of our numerical experiments are shown in Fig.\ref{fig:ZSff_decay}(b,c).
For $M_S<60$GeV, our model predicts a branching ratio around $10^{-8}-10^{-6}$ times of the SM $Br(V\ra f\bar{f})$.
It is interesting that there is a lower bound for these branching ratios and this is understood as the scalar potential stability requires a relatively large $S-H$ mixing.

\begin{figure}[ht]
    \centering

    \subfigure[]{\includegraphics[width=7cm]{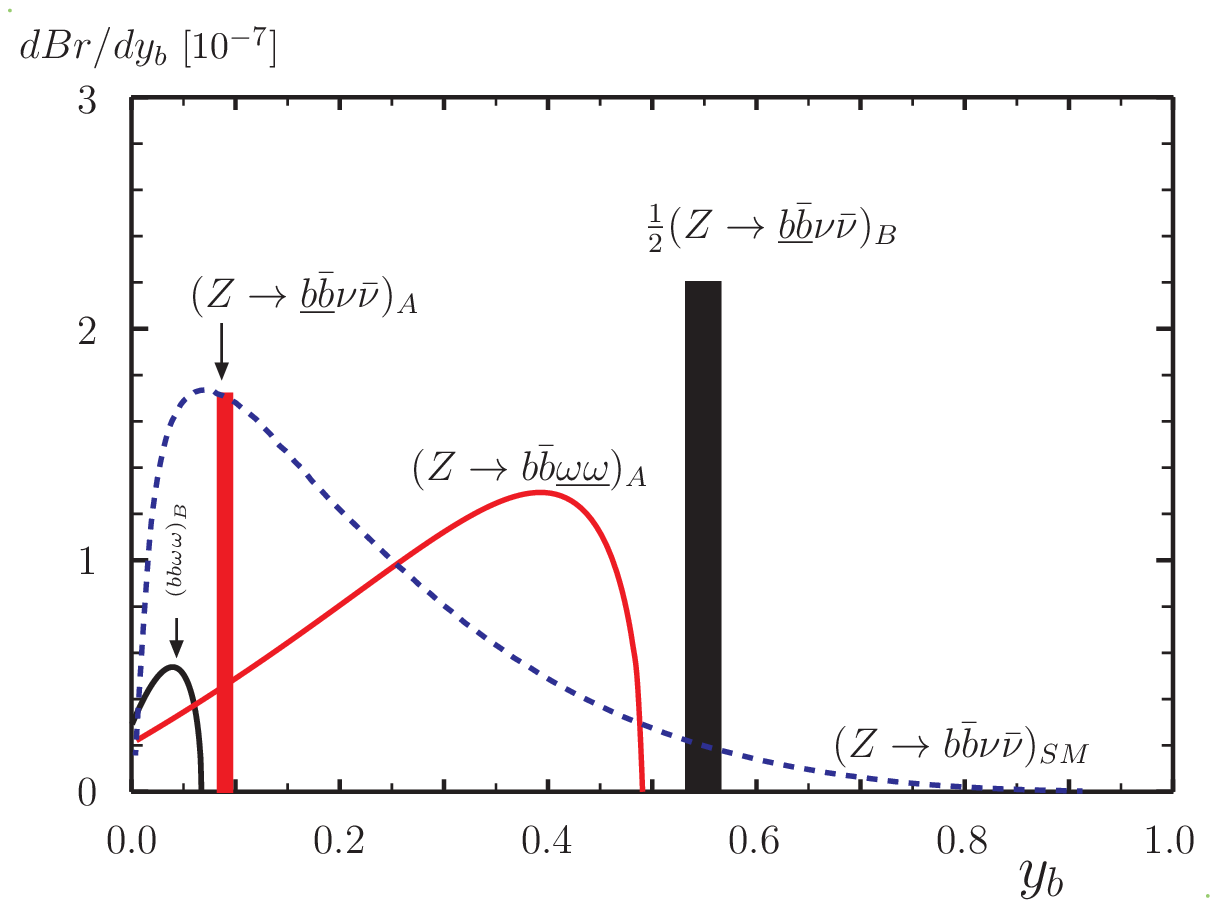}}%
    \qquad
     \subfigure[]{\includegraphics[width=7cm]{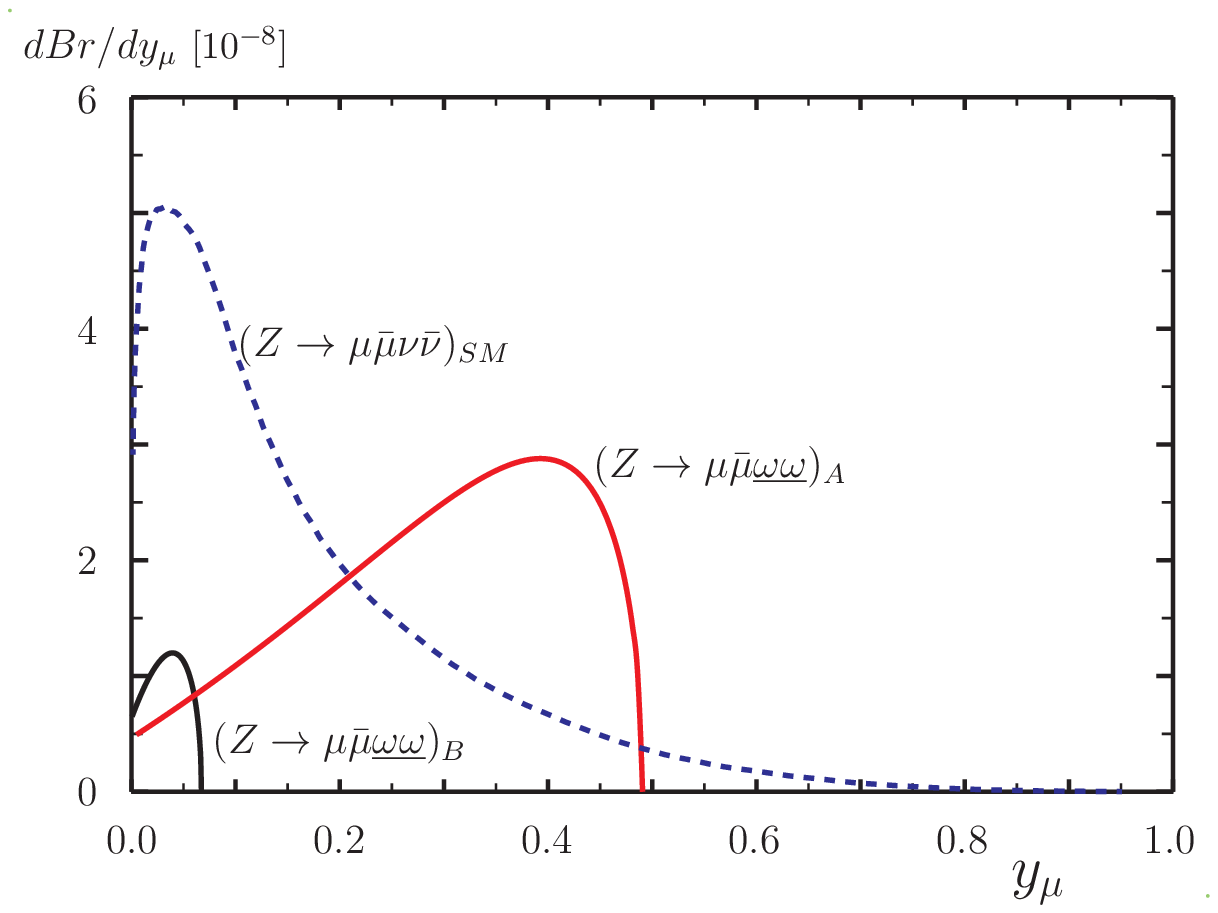}}%
\caption{ (a)Differential branching ratios for Configs.-A and -B v.s. $y_{b}$.
The solid bar represents the decay that $Z\ra S\nu\bar{\nu}$ and then $S\ra b\bar{b}$.
This branching ratio is $2.25\times 10^{-9}(1.43\times 10^{-8})$ for CfA(CfB). The width of the $S$-resonance is much smaller than the
 precision of measuring $m_{b\bar{b}}$ which we take $\pm 1 GeV$ as a bench mark.
 In both panels, the solid curves are the differential branching ratio for $Z\ra S b\bar{b}; S\ra \omega\omega$,
  and  the dashed lines are the SM background.
  (b)Differential branching ratio for Configs.-A and B  v.s. $y_{\mu}$.
  And the tow continuous spectrums are same for $e^+e^-$.
  The branching ratio for $Z\ra S\nu\bar{\nu}; S\ra \mu\bar{\mu}$ are too small due to the muon Yukawa suppression and completely buried in the background. }

    \label{fig:ZSff_diffBr}%
\end{figure}

In order to make the best use of the 3-body decays we note that the dominant branching modes $S$ is into
 $b\bar{b}$ or $\omega \omega$. In turn they lead to the signatures $Z\ra f\bar{f} +b\bar{b}$ and
 $Z\ra f+\bar{f}+ \slashed{E}$ where $\slashed{E}$ denotes missing energy. The particularly interesting ones are $f=b,\mu,e$. The invariant mass squared distribution, $M_{ff}^2$, is a very useful quantity for suppressing the SM background. Defining $y_f=\frac{M_{f\bar{f}}^2}{M_{Z}^2}$  we obtain
 \beqa
\frac{d Br(Z\ra S f\bar{f})}{dy}= \frac{g^2 \sin^2\theta}{192\pi^2\cos^2\theta_W}
\sqrt{y_f^2-2y_f(1+r_Z^2)+(1-r_Z^2)^2}\nonr\\
\times { \left[y_f^2+2y_f(5-r_Z^2)+(1-r_Z^2)^2\right] \over (1-y_f)^2} \times Br(Z\ra f\bar{f})\,,
 \label{eq:invmass}
 \eeqa
 where $r_Z=\frac{M_S}{M_Z}$ and $0\leq y_f \leq (1-r_Z)^2$. The kinematic lower bound can be safely taken to be zero even for $y_b$.
  This distribution peaks at $y$ near the kinematic limit due to the propagator effect since the charged fermion pair comes from a $Z^*$. The dominant SM background
 is due to $Z\ra f^* \bar{f}$ or $\nu^* \bar{\nu}$ follow by the $f^*(\nu^*) \ra f(\nu) +Z^*/W^*/\gamma^*$ with the virtual gauge boson  going into the appropriate final fermions. The $y$ distribution peaks at
 smaller values as seen in Fig.\ref{fig:ZSff_diffBr}. The SM branching ratios for $Z\ra b\bar{b} + \slashed{E}$ is $5.25 \times 10^{-8}$ and for $Z\ra \mu \bar{\mu} +\slashed{E}$ is $1.07\times 10^{-8}$. The latter is a very clean signal to utilize. Furthermore,
 the SM background from $Z\ra Z^* h^*$ is $10^{-4}$ times smaller than the above and can be ignored.

As an illustration we use configurations-A(CfA) and -B(CfB) to bring out the usefulness of the above discussion.
CfA has a relatively small mixing, i.e. $s_\theta^2 =0.00071$, a relatively light $M_S$, and relatively large $Br(S\ra \omega\omega)$.
On the other hand, CfB has a relatively large mixing, i.e. $s_\theta^2 =0.098$, a relatively heavy $M_S$, and relatively small $Br(S\ra \omega\omega)$ but large $Br(S\ra b\bar{b})$. Their corresponding locations in parameter space are marked by the crosses and daggers in Fig.\ref{fig:ZSff_decay} (b-f).
 With an expected $10^{12}$ Z events, CfB(CfA) will have $1.4\times 10^4 (2.3\times 10^3)$ events with $m_{bb}$ peaks at $67.5(27.3)$ GeV.
 And for CfB the signal stands out from the SM background. On the other hand, the continuous $y_b$ distribution for CfA which peaks at around $y_b=0.49$ can be clearly distinguished from the SM background  which peaks at around $y_b\sim 0.07$.
 Similarly, the continuous $y_{\mu,e}$ distribution for CfA also  peaks at around $y_{\mu,e}=0.49$ away from the SM distribution which
peaks at around $y_{\mu,e}\sim 0.05$.

 In passing we also note that
 the decay $Z\ra\omega\omega\nu \bar{\nu}$ will contribute to the $Z$ invisible decays but only at
 level $<10^{-6}$. This will be difficult even at the Z-factory mode of the FCC-ee.

 For $M_S<M_W$ the decay channels $W\ra S+W^*$ will also be open. The virtual $W^*$ will then decay into a fermion  pair. The signals will be similar to the Z decays discussed before. However, we find that it will not add any additional information. It also suffers from lower event rates at the FCC-ee  compared  to the $Z$.

\subsection{DM bound state}
\begin{figure}[ht]
  \centering
  \includegraphics[width=0.36\textwidth]{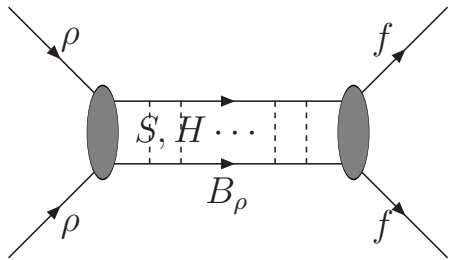}
 \caption{The Feynman diagram for two DM forming a bound state, $B_\rho$, and decays into final state $ff$.}
 \label{fig:BS_annih}
\end{figure}

Our solutions indicate that the DM $\rho$ has mass in the TeV range. Furthermore the parameter
$\lambda_\Phi$ is mach smaller than $\lambda_{\Phi H}$ and $\lambda_{\Phi S} $. For DM with a mass of a few TeV or higher, all the masses  of other states except $\chi$ can be ignored.
More importantly the dark scalar $\rho$ can interact with each other through exchanging the relatively light $S$ and $H$ in the t-channel and this force is attractive. The relevant interaction is given by
\beq
{\cal L} \supset \frac{1}{2} [\lambda_{\Phi H} v_H h + \bar{\kappa} s]\rho^2\,,
\eeq
and  since $\bar{\kappa}\gg \lambda_{\Phi H} v_H$ the $s$ mediation dominates.
As shown in Fig.\ref{fig:lam_Ys}(b), our numerical indicates that $\bar{\kappa}/M_\rho \in [-1.0,1.0]$ and centers around zero. There are considerable number of configurations with both $\bar{\kappa}$ and $M_\rho$ in the range of a few TeV.
In this region of parameter space, two $\rho$'s may form a scalar bound state, $B_\rho$. This possibility can have interesting cosmological consequences as pointed out in \cite{GS}.
Thus, we are led to investigate the  DM-DM annihilation cross section $\langle \sigma v\rangle$ and how this
 quantity may change due to the formation of bound states. For simplicity we will only consider the lowest spin 0 bound state of two $\rho$'s. In the following, we qualitatively discuss  bound state effects in two cases: (i) around the epoch of DM freeze-out where the relative velocity, $v$, between two DM's is relevant for the relic density calculation, and (ii) at present, where $v \ll 1$  and this is important for DM indirect detection.
 The thermal average annihilation cross section due to the $B_\rho$ resonant is schematically represented by the Feynman diagram of Fig.\ref{fig:BS_annih} and it involves three ingredients: (1) the $\rho\rho B_\rho$ coupling vertex, (2) the decay of nearly on-shell $B_\rho$, and (3)  the $B_\rho$ propagator.

We start with the  $\rho\rho B_\rho$ coupling vertex.
If one writes the effective coupling between the bound state $B_\rho$ and $\rho$ as
\beq
 {\cal L} \sim \alpha_{B} B_\rho \rho^2\,.
 \eeq
By dimensional analysis, $\alpha_{B}$  can be estimated to be $ \alpha_{B} \sim (\bar{\kappa}^2/M_\rho) $.

Next, the decay width of  $B_\rho$ is proportional to its wave function absolute squared at the origin
times the decay amplitude squared
$\Gamma_B \propto |\psi(0)|^2 \times |{\cal M}_{B_\rho}|^2$.
 The probability density for two $\rho$'s to meet is $|\psi(0)|^2 \sim \bar{\kappa}^6/M_\rho^3$  by dimension analysis.
We rescale the  decay amplitude square
to make it dimensionless and it can be further broken into
\beq
|{\cal M}_{B_\rho}|^2 = \gamma_{ss}+\gamma_{HH}+\gamma_{sH}+\gamma_{\omega\omega}+\gamma_{W,Z}+\gamma_{f\bar{f}}\,,
\eeq
where the subscripts label the decay final state.
By setting all final states massless, and  with the help of Eq.(\ref{eq:sigv}), we immediately have
\footnote{Note that if $M_S=M_H$ there is no way to distinguish these two neutral scalars and the mass basis and interaction basis can be made equal or $\theta=0$ effectively. }
\beqa
\gamma_{ss} \simeq \left[\lambda_{\Phi S} - \frac{\bar{\kappa}^2}{M_\rho^2}\right]^2\,,\,\,
\gamma_{HH} \simeq \lambda_{\Phi H}^2\,, \nonr\\
\gamma_{\omega\omega} \simeq \left[\lambda_{\Phi S} -  \frac{ \kappa ^2 }{M_\rho^2 -\kappa v_S } \right]^2\,,\,\,
\gamma_{W,Z}\simeq 3 \lambda_{\Phi H}^2\,,
\eeqa
where we have dropped terms suppressed by ${\cal O}(v_H /M_\rho)$.
Since both $\gamma_{sH}={\cal O}(v_H^2/M_\rho^2) $  and $\gamma_{f\bar{f}}= {\cal O}(m_f^2/M_\rho^2 )$ thus can be neglected, and
\beq
|{\cal M}_{B_\rho}|^2 \simeq \gamma_{ss}+ \gamma_{\omega\omega}+ 4 \lambda_{\Phi H}^2\,.
\eeq
The dimensionless factor $|{\cal M}_{B_\rho}|^2$ in our numerical analysis is $\sim {\cal O}(1)$ and the bound state decay width can be estimated :
\beq
\Gamma_B \sim M_\rho \left(\frac{\bar{\kappa}}{M_\rho}\right)^6 [\gamma_{ss}+ \gamma_{\omega\omega}+ 4 \lambda_{\Phi H}^2 ]\,.
\eeq

Finally, we put $\Gamma_B$ into the propagator squared and the annihilation cross section due to the $B_\rho$ resonant
can be estimated to be
\beq
 \sigma v  \sim {\alpha_B^2  ( \Gamma_B/M_B ) \over (s-M_B^2)^2+\Gamma_B^2 M_B^2}\,,
 \label{eq:sigmav_BS_BOE}
\eeq
where the factor $\Gamma_B/M_B$ is inserted to take care the nearly on-shell $B_\rho$ decay.

When $v \ll 1$, $s\sim M_B^2$ and there is almost no temperature dependence, we have
\beq
\langle \sigma v \rangle \sim {\alpha_B^2  ( \Gamma_B/M_B ) \over (s-M_B^2)^2+\Gamma_B^2 M_B^2}\sim  \frac{\alpha_B^2 }{M_B^3 \Gamma_B} \sim  \frac{R_B }{M_\rho^2 }[\gamma_{ss}+ \gamma_{\omega\omega}+ 4 \lambda_{\Phi H}^2 ]\,,
\eeq
  and
\beq
R_B\equiv  \left(\frac{M_\rho}{\bar{\kappa}}\right)^2 [\gamma_{ss}+ \gamma_{\omega\omega}+ 4 \lambda_{\Phi H}^2 ]^{-2}
\eeq
is the boost factor for indirect DM detection.

For a typical value that $|{\cal M}_{B_\rho}|^2 \sim {\cal O}(10^0)$ and $\bar{\kappa}\sim 0.1 M_\rho$ we have the boost factor around $100$. In our numerical study, we found that the branching ratio of DM pair annihilate into mono-energetic Majoron pair is  a few to $40\%$, Fig.\ref{fig:DM_ann_BR_listplot}(c).  The boost factor will make $\langle\sigma v\rangle(DM+DM\ra \omega\omega) \sim 10^{-26} -10^{-24}(cm^3/s)$.
 The sizable annihilation cross section  opens up a possibility   that  the Goldstone bosons could be a component of the `apparent' neutrino flux at $E_\nu= M_\rho$ in IceCube and other neutrino observatories. Moreover, they give rise to shower events and no tracks and it is mostly originated from the Galactic center.
In additional to the mono-energetic Majoron line, there is sizable fraction that DM pair annihilate into $SS$ pair, see Fig.\ref{fig:DM_ann_BR_listplot}(b). The mono-energetic $S$ then subsequently decays into $b\bar{b}$ and $\omega\omega$.
Our numerical experiment indicates that $Br(S\ra \omega\omega) > Br(S\ra b \bar{b})$ in most of the parameter space, Fig.\ref{fig:S_decay}. This secondary Majoron  contributes a continuous spectrum with a peak at $E_\omega = M_\rho/2$ and a total cross section about twice of that of the Majoron line.
This continuous Majoron spectrum  completely overlaps with the neutrino spectrum from $DM+DM\ra ZZ, WW$ and $Z, W$ subsequently decay into neutrinos. An  immediate prediction is that the shower/track ratio in the continuous `apparent' neutrino spectrum is larger than the SM one and the line gives shower-like events. The details will be left for further studies.

Now, we estimate the thermal average annihilation cross section around the DM freeze-out temperature, $T_\rho$,
which typically takes a value $T_\rho \sim M_\rho/20$ and thus $s\sim ( M_B^2 +4 T_\rho^2)$.
It is required that $\bar{\kappa} \lesssim 0.4 M_\rho$ such that the BS width term in the denominator of Eq.(\ref{eq:sigmav_BS_BOE}) is less important than the $T_\rho$ contribution. Therefore the $\langle \sigma v\rangle$ becomes:
\beq
\langle \sigma v \rangle \lesssim {\alpha_B^2  ( \Gamma_B/M_B ) \over 16 T_\rho^4}\,.
\eeq
Comparing to the  $\langle \sigma v\rangle_0$ without BS, which is mainly controlled by Eq.(\ref{eq:sigv}),
we gain an enhancement factor about
\beq
\sim \frac{M_\rho^4 }{32 T_\rho^4} \left(\frac{\bar{\kappa}}{M_\rho}\right)^{10} \lesssim 1
\eeq
for   $T_\rho \sim M_\rho/20$ and $\bar{\kappa} \lesssim 0.4 M_\rho$.
So that the bound state effect at the DM freeze out era is not important comparing to the direct DM-DM tree-level annihilation in our model.

\subsection{Kinetic decoupling between DM and DR}
Even after the thermal decoupling between the DM($\rho$) and DR ($\omega$), they can still interact with each other through the scalar quartic coupling term $\frac14 \lambda_{\Phi S} \omega^2 \rho^2$.
Given that $\lambda_{\Phi S}$ is sizable in this model, we would like to know whether  this has any detectable cosmological implication. And the relevant question to ask is at what temperature, $T_k$,  the two will decouple kinetically.

 By straightforward calculation, one has the nonrelativistic cross section for $\omega + \rho \ra \omega + \rho$
\beq
\sigma_{\omega \rho \ra \omega  \rho} ={|\lambda_{\Phi S}|^2 \over 32 \pi M_\rho^2}
\eeq
for  Majoron energy  is much less than $M_\rho$.
After taking the thermal average, the rate for a single DM particle to collide with a Majoron is given by
\beq
\Gamma_{col}\equiv \left\langle n_\omega  \sigma_{\omega \rho \ra \omega  \rho} v \right\rangle  = \frac{\zeta(3)}{\pi^2}{|\lambda_{\Phi S}|^2 \over 32 \pi M_\rho^2} T^3\,,
\eeq
where $n_\omega$ is the DR number density which behaves like that of photon and $T$ is the temperature
\footnote{ And we ignore the difference between  $T_\omega$ and the photon temperature in this order of magnitude estimate. }.
At low temperatures, the typical momentum of $\omega$,  $p_\omega\sim{\cal O}(T)$,  is much less than the typical
momentum of DM, $p_\rho \sim {\cal O}(\sqrt{M_\rho T})$. Therefore, for a DM to acquire a momentum transfer which is comparable to $p_\rho$, it needs to accumulate many tiny momentum transfers from  multiple collisions with the ambient DR. This process is very similar to the random walk and the number of collisions can be estimated to be $\sqrt{N_{col}} p_\omega \sim p_\rho $ or $N_{col}\sim M_\rho/T $.
And the kinetic decoupling temperature can be estimated by requiring that
\beq
\frac{\Gamma_{col}(T_k)}{N_{col}(T_k)} \simeq H(T_k) \simeq \frac{T_k^{2}}{ M_{pl}}\,,
\eeq
or
\beq
T_k \sim \left( { 32 \pi^3 M_\rho^3 \over \zeta(3) |\lambda_{\Phi S}|^2 M_{pl}}\right)^{\frac12}\,.
\eeq
Using the above estimate, we obtain the corresponding $T_k= 0.67 (2.34)$ MeV for configuration-A(B). Overall,  the kinetic decoupling between DM and DR happens at around ${\cal O}(0.1)- {\cal O}(1)$ MeV in our model.
Above $T_k$, DM and DR form a tightly bounded fluid. When the DM gravitate due to the positive density fluctuation, the compressed DR provides a resilient pressure. And the resulting acoustic oscillation erases the small scale density perturbation. Thus, the temperature $T_k$ determines a lower bound on the masses of the smallest halos from the Jeans mass\cite{Loeb:2005pm},
\beq
M_{cut} \sim 10^{-4} \left( {10 \mbox{MeV} \over T_k}\right)^3 M_\odot\,.
\eeq
Currently, the highest kinetic decoupling temperature can be probed is around $10$keV,
and the $T_k$ in our model is too high to be detected  with the current observational precision.

\section{Conclusion}
We calculated the 1-loop beta functions for the minimal singlet Majoronic model\cite{CN} and performed a thorough
numerical study on the parameter space of this model.
In order to have an operational type-I see-saw mechanism, it is required that the lepton number breaking scale, $v_S$, is lower than
the scale $\mu_{VS}$ where the SM electroweak vacuum become unstable.
The extra scalar degrees of freedom always help to improve the stability of  SM electroweak vacuum, thus $\mu_{VS}>\mu_{VS}^{SM}$.
However, the right-handed Majorana neutrinos contribute negatively to the beta function for $\lambda_S$ through the Yukawa $Y_S$. Additional attention to this new instability has to be taking into account and ensure that $\lambda_S >0$ when energy scale $\mu < \mu_{VS}$.
 Moreover, the beta function for $\lambda_\Phi$ is always positive so we looked for the solutions that there is no Landau pole below $\mu_{VS}^{SM(1-loop)}$.
Other phenomenological requirements had been considered in our numerical scan are: (1) the upper limit of SM Higgs invisible decay width, (2) Majoron decouple from the thermal bath at the temperature between $m_\mu$ and 2 GeV, (3) the upper limit on the mixing between the SM Higgs and the beyond SM scalar, (4) the correct DM relic density, (5) the upper limit of direct DM searches.

The results of our numerical experiment have been summarized and discussed in Sec.III. Here we highlight the physics of our finding.
\begin{enumerate}
\item A decoupling temperature at or below $2GeV$ leads to small $\lambda_{SH}$, Eq.(\ref{eq:dec}).
\item For the sake of $\lambda_S$-vacuum stability, Eq.(\ref{eq:RGE}c), $y_S =\sqrt{2}M_N/v_S$ cannot be too large. This leads to $v_S$ in the $2-20$ TeV range which is relative large compared to the SM VEV.
\item In order to have such a value for $v_S$, a large mixing angle between $S$ and the SM Higgs is preferred, Eq.(\ref{eq:VS_mass_basis}).

 \item From the direct search for the light neutral scalar, a large mixing angle is only permitted when $M_S$ is in the range of $10-100$ GeV and a higher $T_{dec}$ follows, Fig.\ref{fig:MS_max_mixing}.
\item To counteract the negative contribution from $y_S$, a sizable and positive $\lambda_{\Phi S}$  is needed, Eq.(\ref{eq:RGE}c).

\item To improve SM vacuum stability, sizable $\lambda_{\Phi H}\sim {\cal O}(1)$ is needed,  Eq.(\ref{eq:RGE}a).
\item $\lambda_{\Phi H}\sim {\cal O}(1)$  leads to heavy $M_\rho>1.5$ TeV to keep the thermal average cross section $\langle \sigma v\rangle$ under $2.5\times 10^{-9}(GeV)^{-2}$ at the freeze-out .
\item The RGE of $\lambda_\Phi$ prefers small $\lambda_\Phi$ and $\lambda_{\Phi S}$ to avoid the Landau pole below $\mu_{VS}$.
\item Since $\lambda_{\Phi S}$ cannot grow indefinitely, a DM with $M_\rho>4$ TeV will yield too small $\langle \sigma v\rangle$
and render too much relic density.
\end{enumerate}

Phenomenologically, this model predicts a universal signal strength $\mu_i=\cos^2\theta$ and the parameter
space with $M_S\gtrsim 40$ GeV can be probed indirectly at the LHC with $3ab^{-1}$ luminosity.
If not excluded by LHC14, the triple-Higgs coupling in the same parameter space can be further tested at the ILC, CLIP, or VHC.
 Additional signatures can also be searched for in the Z-factory mode of FCC-ee. The decays
of $Z\ra b\bar{b}+\slashed{E}$ is particularly sensitive to the existence of a light $S$ which mixes
with the SM Higgs boson. Using the invariant mass distribution of the b pairs one can probe
mixings $\sin^2\theta \lesssim 10^{-3}$.
Finally, the DM bound state could yield a boost factor around $\sim 100$. And $\rho+\rho\ra \omega+\omega$ annihilation at the galactic center will generate  shower events with an apparent neutrino energy $E_\nu=M_\rho$ in IceCube and other astronomical neutrino observatories.

\begin{acknowledgments}
 We thank Florian Staub for pointing out the terms in the beta function we omitted in an earlier version.
 WFC was supported by the Taiwan MOST under
Grant No.\ 102-2112-M-007-014-MY3. J.N.N is partially supported by the NSERC and National Research Council of Canada through a contribution to TRIUMF.
\end{acknowledgments}

\end{document}